\documentclass[final,authoryear,5p,times,twocolumn]{elsarticle}

\usepackage{amssymb}

\journal{Icarus}

\begin{document}

\begin{frontmatter}



\title{
3D modelling of the early Martian Climate under a denser CO$_2$
atmosphere: Temperatures and CO$_2$ ice clouds.
}


\author[lmd]{F. Forget\corref{cor1}}
\author[lmd]{R. Wordsworth}
\author[lmd]{E. Millour}
\author[lmd]{J-B. Madeleine}
\author[lmd]{L. Kerber}
\author[lmd]{J. Leconte}
\author[latmos]{E. Marcq}
\author[arc]{R.M. Haberle}

\cortext[cor1]{Corresponding author. E-mail: forget@lmd.jussieu.fr}

\address[lmd]{LMD, Institut Pierre-Simon
Laplace, Université P. et M. Curie BP99, 75005 Paris , France}
\address[latmos]{LATMOS, Institut Pierre-Simon Laplace,
78280 Guyancourt, France}
\address[arc]{NASA Ames Research Center, Space Science Division, MS 245-3,
Moffett Field, CA, 94035-1000, USA.}

\begin{abstract}

On the basis of geological evidence, it is often
stated that the early martian climate was warm enough for liquid water to flow
on the surface thanks to the greenhouse effect of a thick atmosphere.
We present 3D global climate simulations of the early martian climate performed
assuming a faint young sun and a CO$_2$ atmosphere with surface pressure between
0.1 and 7 bars.
The model
includes a detailed representation of radiative transfer using
revised CO$_2$ gas collision induced absorption properties,
and a parameterisation of CO$2$ ice cloud microphysical and radiative
properties. A wide range of possible climates is explored using various
values of obliquities, orbital parameters,
cloud microphysic parameters, atmospheric dust loading, and surface properties.

Unlike on present-day Mars, for pressures higher than a fraction of a bar 
surface temperatures vary with altitude because of adiabatic cooling /
warming of the atmosphere.
In most simulations, CO$_2$ ice clouds cover a major part of the planet.
Previous studies suggested that they could have warmed the
planet thanks to their scattering greenhouse effect.
However, even assuming parameters that maximize this effect,
it does not exceed +15~K.
Combined with the revised CO$_2$ spectroscopy and the impact of
surface CO$_2$ ice on the planetary albedo,
we find that a CO$_2$ atmosphere could not have raised the
annual mean temperature above 0$^\circ$C anywhere on the planet.
The collapse of the atmosphere into permanent
CO$_2$ ice caps is predicted for pressures higher than
3~bar, or conversely at pressure lower than one bar if the obliquity
is low enough.
Summertime diurnal mean surface temperatures above 0$^\circ$C
(a condition which could have allowed rivers and lakes to form)
are predicted for obliquity larger
than 40$^\circ$ at high latitudes but not in locations
where most valley networks or layered sedimentary units are observed.
In the absence of other warming mechanisms,
our climate model results are thus consistent with a cold
early Mars scenario in which non climatic mechanisms must occur to explain
the evidence for liquid water.
In a companion paper by Wordsworth et al., we simulate the
hydrological cycle on such a planet and discuss how this could
have happened in more detail.

\end{abstract}


\end{frontmatter}




\section{Introduction}

Spacecraft sent to Mars in recent years have revealed new
evidence suggesting that the environmental conditions on
early Mars were very different than those today,
with liquid water flowing on the surface at least episodically.
Geomorphological evidence of past water flow in
ancient terrains includes
valley networks (remnant
of widespread fluvial activity) (Carr, 1996), extensive sedimentary
layered deposits (including delta-like landforms)(Malin and Edgett, 2000, 2003)
and erosion rate much higher than today affecting the most ancient
landforms  (Craddock and Maxwell, 1993).  Satellite remote sensing and
in-situ analysis of the surface mineralogy have also revealed
the local presence of minerals that require
liquid water for their formation: clay/phyllosilicates
(Poulet et al. 2005, Bibring et al. 2006, Ehlmann et al. 2011), sulfates
(Gendrin et al. 2005, Squyres et al. 2004),
opaline silica (Squyres et al. 2008), carbonate (Ehlmann et al. 2008,
Boynton et al. 2009, Morris et al. 2010) and chloride (Osterloo et al., 2008).

\nocite{Carr:96,Mali:00lay,Mali:03,Crad:93,Poul:05,Ehlm:11,Gend:05}
\nocite{Squy:04,Squy:08,Ehlm:08,Boyn:09,Morr:10,Oste:08}

All these observations have provided a wealth of information
about early Mars.  In particular, they suggest that the conditions
strongly varied throughout the Noachian and Hesperian era.
Nevertheless, we still do not know if the conditions suitable for
liquid water were stable on long timescales, or if they were
the consequence of episodic, possibly catastrophic events. What is
clear is that the conditions which allowed such extensive alteration by
liquid water occurred early in the history of Mars and not later.
In particular, the fact that some alteration minerals can still be observed
today, although they formed several billion years ago and
are easily transformed into other materials by, e.g., diagenesis,
suggests that water has been very limited on the martian surface
from soon after their formation until today (Tosca and Knoll, 2009)).
Similarly the majority of the  valley networks are found almost
exclusively in the highly cratered ancient southern highlands,
which date back from the end of the heavy bombardment period
some 3.5-3.8 Gyr ago (Fasset and Head, 2008).
\nocite{Tosc:09,Fass:08}

Although the role
of hydrothermalism (resulting from volcanism or impacts) with
respect to atmospheric induced processes is not clear, climate
conditions were certainly different on early Mars.
It is possible that
the atmosphere could have been thicker than today, providing an environment
more suitable for liquid water thanks to a surface pressure
well above the triple point of water, and possibly a greenhouse effect
capable of warming
the surface closer to the melting points at 0$^{\circ}$C.
In analogy with Earth and Venus, even after taking into account
the differences in size, it can be estimated that the initial inventory
of volatiles on  Mars included at least several bars of atmospheric
gases, mostly CO$_2$ (see Haberle et al. 1998). On this basis, since the
1980s many studies have been performed to characterize the possible
climate on early Mars assuming a CO$_2$ atmosphere thicker than today,
and taking into account the fact that according to stellar evolution
models the Sun's
luminosity at 3.8 Ga was 75\% of its present value
(e.g., Gough, 1981).
\nocite{Habe:98,Goug:81}
Almost all of these studies were performed using one-dimensional (1-D)
radiative
convective models (see Section~\ref{sc:previous}).
Characterization has been challenging however, because of the complex
CO$_2$ gas spectroscopy, the likely formation of CO$_2$ ice clouds in the
atmosphere, and the four-dimensional aspects of planetary climate.
The primary goal of the present
paper is to update these calculations using a full
3-D climate model including a parameterization of CO$_2$ ice clouds and state
of the art spectroscopic data. In a companion paper (Wordsworth et al., 2012)
we extend these calculations by including a model of the possible
water cycle that takes into account the radiative effects of water vapor and
clouds, and can predict precipitation
and the formation of lakes and glaciers on early Mars.
To start with, in Section~\ref{sc:previous} and \ref{sc:which_atm}
 we review previous
work on modelling early Mars climate, and briefly discuss the possible early
Mars atmosphere composition and thickness.
Section~\ref{sc:model} describes
our Global Climate Model (GCM). The simulated climates for various pressures,
cloud microphysics parameters, obliquity values,
orbital parameters, and possible
atmospheric dust loading scenarios are analysed in
Section~\ref{sc:results}.
Finally, in Section~\ref{sc:discussion}, we discuss and summarise our results.

\nocite{Word:12}

\section{Previous Modelling Studies}

\label{sc:previous}

The climate modelling studies of early Mars performed before 1998 are reviewed in
detail in Haberle~(1998). By the end of the 1980s, the
paradigm for early
Martian climate was based on the results of Pollack et al. (1987). \nocite{Poll:87}
Using a
1-D radiative convective model, they had shown that a 5-bar gaseous CO$_2$ atmosphere
would raise the global mean surface temperature to 0$^{\circ}$C, allowing ``warm and wet'' conditions
(though large, 5~bars of CO$_2$ was still consistent with contempory
estimates of the available
inventory). This scenario was later challenged when Kasting (1991) published his reanalysis
of the Pollack et al. (1987) greenhouse calculations.
Kasting used the same model, but took
into account the fact that at higher pressures CO$_2$ can condense
in the middle atmosphere.
\nocite{Kast:91,Poll:87}
This seriously decreases the greenhouse effect because of the latent heat
warming of the middle atmosphere. Furthermore Kasting suggested that the
resulting CO$_2$ ice clouds would probably raise the planetary albedo and further cool
the planet. However, still working with the same model, Forget and Pierrehumbert (1997)
later showed that the inclusion of the
CO$_2$ ice cloud radiative effect at both solar and thermal
infrared wavelengths led to a
net warming of the surface rather than a cooling. Indeed, assuming that the CO$_2$
ice cloud particles are larger than a few micrometers,
they can readily scatter infrared radiation and reflect
outgoing thermal radiation back to the surface. The resulting warming
effect more than compensates for the albedo increase due to the clouds.
These calculations were
later confirmed by Mischna et al. (2000), who used a more sophisticated and accurate
1-D radiative-convective transfer model based on correlated k-distribution methods.
They noted that
CO$_2$ ice clouds could also cool the surface if they are low and optically thick,
and concluded that ``estimating the actual effect of CO$_2$ clouds on
early martian climate
will require three-dimensional models in which cloud location,
height, and optical depth, as
well as surface temperature and pressure, are determined self-consistently''.
Nevertheless, the physics of carbon dioxide clouds in a dense
CO$_2$ atmosphere was further
explored by Colaprete and Toon (2003) still using a 1-D atmospheric model,
but including a detailed microphysical cloud model.
They took into account laboratory studies by Glandorf
et al.  (2002) showing that high critical supersaturations are required for CO$_2$
cloud particle nucleation and that surface kinetic growth is not limited. Under such conditions,
Colaprete and Toon (2003) predicted large carbon dioxide ice particles with radii greater
than 500 $\mu$m, and thus low cloud opacities. Because of this, and because of the
warming of the atmosphere associated with cloud formation, they estimated that the greenhouse
effect of CO$_2$ clouds would probably be limited to 5-10 K surface warming.
\nocite{Forg:97,Misc:00,Cola:03a,Glan:02}

All these studies took into account the radiative effect of CO$_2$
and water vapor.
Postawko and Kuhn (1986) also explored the possible greenhouse effect
of SO$_2$. Yung et al. (1997) resumed this investigation, and
studied the radiative effect of a very small amount (0.1 ppmv)
of SO$_2$ in a 2 bar CO$_2$ atmosphere in a 1D model. They showed that it would
raise the temperature of the middle atmosphere by approximately 10~K, so
that the upper atmosphere
would globally remain above the condensation temperature of
CO$_2$. Exploration of the impact of SO$_2$ was further motivated by the discovery of
sulfate sediments in Mars' ancient terrains (Gendrin et al. 2005, Squyres et al. 2005,
Halevy et al. 2007).
Johnson et al. (2008) investigated the impact of larger amount of
sulfur volatiles (H$_2$S and SO$_2$ mixing ratios of 1 to 1000 ppmv) in a martian
atmosphere of 50 and 500 mbar of CO$_2$ (with H$_2$O). For this purpose they used
the 3D Mars Weather Research and Forecasting (MarsWRF) GCM
(Richardson et al., 2007).
Their control simulations, achieved with a pure CO$_2$ atmosphere or
with CO$_2$ + H$_2$O (but neglecting the radiative effects of
CO$_2$ clouds or H$_2$O clouds)
are of interest in the present paper for comparison.
Calculations performed including the sulfur volatile influxes suggested that
these gases could have been responsible for greenhouse warming up to 25 K above
that caused by CO$_2$.  However, Tian et al. (2010) showed that
such large amounts of SO$_2$
would inevitably lead to the formation of sulfate and large reduced sulfur
aerosols (S$_8$). They concluded
that these aerosols would have raised the planetary albedo and that the
resulting cooling  would
 more than outweigh the gaseous greenhouse effect.
\nocite{Post:86,Yung:97,Gend:05,Squy:05,Hale:07,John:08,Rich:07,Tian:10}

In this paper, we do not include the effect of sulfur volatiles or aerosols.
Our goal is to investigate the details of the climate induced by a CO$_2$ atmosphere
as simulated by a 3D GCM in which CO$_2$ cloud location, height,
and optical depth, as well as surface temperature and pressure, are
determined self-consistently for the first time.
In comparison to the previous studies mentioned
above, we also benefit from an improved parameterization of the collision-induced
absorption of CO$_2$ (important for pressure larger than a fraction of a bar)
that we have developed for this project, as described in
Wordsworth et al. (2010) (see details in Section~\ref{sc:cia}).
To calculate the collision-induced opacity in a CO$_2$ atmosphere,
all the studies mentioned above relied without modification on a
parameterization originally derived for the Venus atmosphere by James
Pollack (Pollack et al., 1980; see detailed description in Kasting et al. 1984).
It was based on the measurements of Ho et al.
(1971) from 7 to 250~cm$^{-1}$, and on a simple parameterization of collision
induced opacity in the other spectral domains described in an unpublished
PhD thesis by John F. Moore (Moore ,1971), still available as a NASA report).
This parameterization included significant opacities between 526 and
295~cm$^{-1}$ resulting from the pressure induced wings of the strong 15~$\mu$m
bands. Unfortunately, this feature was kept in subsequent models, despite
the fact that these opacities were probably overestimated and, in most cases,
already accounted for in the codes chosen to calculate the radiative transfer
in the 15~$\mu$m  band. As a result, most  previous studies probably
overestimated the greenhouse warming of CO$_2$ by several Kelvins.
\nocite{Word:10co2,Poll:80jgr,Kast:84,Ho:71,Moor:71}

\section{Which atmosphere for early Mars?}

\label{sc:which_atm}

By scaling the Earth or Venus volatile inventory,
one can estimate that the amount of CO$_2$ brought to Mars
during accretion
 was probably larger than 10~bars. However, Tian et al.
(2009) showed that the extreme ultraviolet flux from the young sun was so high
that it would have strongly warmed the thermosphere and induced thermal escape
of a primordial  CO$_2$-dominated martian atmosphere.
In their calculation, a CO$_2$ atmosphere
could not have been maintained until about 4.1 billion years ago. Under such
conditions,
how much atmosphere could have been degassed late enough after 4.1 Ga
during the mid to late Noachian?
Phillips et al. (2001) estimated that the total release of gases from
the magmas that formed Tharsis during the Noachian era could have produced
the integrated equivalent of a 1.5-bar CO$_2$ atmosphere. However, they assumed
a magmatic CO$_2$ content of 0.65 weight percent, consistent with Hawaiian
basaltic lavas. Since then, several authors have suggested that this was
probably overestimated. In particular, Hirschmann and Withers (2008)
calculated that post-4.5 Ga
magmatism could have provided only 0.1 to at most 1 bar of CO$_2$.
To refine these conclusions, Grott et al.  (2011)
combined the  Hirschmann and Withers (2008) model for the solubility of CO$_2$
with a thermo-chemical evolution model to self consistently calculate the
dissolved amount of CO$_2$ in martian magmas. They estimated that during
Noachian, about 250~mbar of CO$_2$ were outgassed between 4.1 and 3.7 Ga.

\nocite{Tian:09,Phil:01,Hirs:08,Grot:11}

In spite of these studies, if one assumes that more than one bar of atmospheric
CO$_2$ was present during the Noachian era, where did it go?
Estimating the amount of atmosphere that could have escaped
to space in the last 4~Gyr is difficult because many
different processes may have been involved, including
photochemical escape,
ions dragged by the solar wind field, sputtering, and 
impact erosion (Chassefiere and Leblanc, 2004).
Up to now, only the present-day Mars ion loss
by solar wind interaction has been measured, and
found to be very small
(Barabash et al., 2007).
The possible loss of volatiles resulting from large impacts (Melosh and
Vickery 1989) is also difficult to model and constrain, but
recent studies suggest that it must have been small
(Pham et al., 2011).
Altogether,
the most recent estimations of the amount of CO$_2$ lost to space in the past
3.5~Gyr are below a few hundreds of millibars
(Leblanc and Johnson, 2002, Lammer et al. 2008, 2010).

\nocite{Bara:07,Chas:04,Melo:89,Pham:11,Lamm:08,Lamm:10epsc}

Alternatively, one classical
hypothesis is to assume that, as on Earth, large amounts of CO$_2$ could be
stored in form of carbonates in the martian crust after chemical precipitation.
However, almost no carbonates were initially detected by the OMEGA imaging
spectrometer in spite of its high sensitivity to the spectral
signature of carbonates (Bibring et al., 2005).
Recently, several observations from orbiters (Ehlmann et al., 2008) and landers
(Boynton et al. 2009,
Morris et al. 2010) have revived the carbonate hypothesis
and reasserted the importance of carbon dioxide in martian climate history
(Harvey, 2010).
\nocite{Bibr:05,Ehlm:08,Boyn:09,Morr:10,Harv:10}


In this paper, we have chosen to explore the possible climate on early
Mars for a wide range of surface pressures,
up to 7~bars, as assumed  in previous works on the same topic.
However, when interpreting our model results, it is important to keep in mind
that atmospheres thicker than one bar may be unlikely.

Another issue for an early Mars CO$_2$ atmosphere is its photochemical
stability. Using a 1-D photochemical model of the martian atmosphere, Zahnle et
al. (2008) showed that in a thick, cold and dry CO$_2$ atmosphere in which a
surface sink is assumed for reactive oxidized gases (like H$_2$O$_2$ and
O$_3$), CO$_2$ would tend to be reduced into CO.
They noted that the process is very
slow, and that ``CO$_2$ atmospheres can be unstable
but persistent
simply because there isn't time enough to destroy them''.

Finally, in this work we did not take into account the presence of
other possible greenhouse gases. Such gases are not expected to be
photochemically stable since they should have been
photodissociated or oxydized, but they  may be present if geochemical
sources were active enough. This is further discussed in
section~\ref{sc:othergases}.

\nocite{Zahn:08}

\section{Global Climate Model description}
\label{sc:model}

\subsection{Generalities}

We have used a new ``generic" version of the
LMD Global Climate Model
recently developed to simulate any kind of atmosphere with the goal of
studying early climates in the solar system (this paper) as well as
possible climates on extrasolar planets (e.g. Wordsworth et al. 2011b).
In practice, the model
is derived from the LMD present-day Mars GCM
(Forget et al. 1999), with several new parameterizations
(see below and Wordsworth et al. (2012)).
This Mars GCM has been used successfully to simulate
Mars meteorology (e.g. Forget et al. 1998,
Montmessin et al. 2004, Madeleine et al. 2011)  and photochemistry
(Lefevre et al. 2004, 2008) from the surface
to the thermosphere, and to simulate recent climate
changes induced by
the oscillations of Mars' rotational and orbital parameters
(Levrard et al. 2004, Forget et al. 2006,
Montmessin et al. 2007a, Madeleine et al. 2009).
The LMD GCM solves the primitive equations of meteorology using a
finite difference dynamical core on an Arakawa C grid.
This dynamical core has been used and tested successfully in many kind of
atmospheres thicker than present-day Mars such as the Earth
(e.g. Hourdin et al. 2004), Venus (Lebonnois et al. 2010), and Titan (Hourdin et
al. 1995, Lebonnois et al. 2012).

\nocite{Word:11ajl,Forg:99,Word:12,Forg:98,Mont:04,Lefe:04,Lefe:08}
\nocite{Levr:04,Forg:06,Mont:07cap,Made:11,Hour:04,Lebo:10,Hour:95,Lebo:12}
\nocite{Word:11mamo,Made:09}

In this paper, simulations were performed with two horizontal resolutions: 32$\times$24
(corresponding to resolutions of 7.5$^{\circ}$ latitude by 11.25$^{\circ}$ longitude)
and 64$\times$48 (3.75$^{\circ}$ latitude by 5.625$^{\circ}$ longitude).
In the vertical, the model uses
hybrid coordinates, that is, a terrain-following
$\sigma$~coordinate system
near the surface and lower atmosphere
($\sigma$ is pressure divided by surface pressure), and pressure levels
in the upper atmosphere. In this work, with the exception of one sensitivity study
described in section~\ref{sc:cloud},
we used 15 layers with the lowest mid-layer levels at about 18~m, 60~m,
150~m, 330~m,
640~m etc., and the top level
at 0.3\% of the surface pressure, that is about 5.3 scale heights
($>$50~km) above the zero datum.

Nonlinear interactions between explicitly resolved scales
and subgrid-scale processes are
parameterized by applying a scale-selective horizontal
dissipation operator based on an $n$ time iterated Laplacian
$\Delta^{n}$.  This can be written
${\partial q}/{\partial t} = ([-1]^{n}/ {\tau_{\mbox{\scriptsize diss}}})
(\delta x)^{2n} \Delta^{n} q$
where $\delta x$ is the smallest horizontal distance represented in the
model, $\tau_{\mbox{\scriptsize diss}}$ is the dissipation timescale for a
structure of scale $\delta x$, and $q$ a variable like temperature, meridional wind,
or zonal wind.
As in most GCMs, this dissipation operator is necessary
to ensure the numerical stability of the dynamical
core. However, it must be used with moderation (i.e. the dissipation
timescales must be kept as high as possible).
In particular, during this study we found that, when used with
large topography variations and with atmospheric pressure larger than about one
bar, our dissipation scheme tended to artificially produce some heat in
the lower atmosphere. This production is usually completely negligible,
but with high pressure and a strong greenhouse effect like in some of
our simulations we found that the
impact on surface temperatures could be non-negligible if the dissipation
timescale were chosen smaller than necessary.
This meant that our initial results overestimated the warming possible due to
CO$_2$ clouds in the early martian atmosphere, and hence the likelihood of warm,
wet conditions under a pure CO$_2$ atmosphere
(Wordsworth et al. 2011a). Fortunately, we
have been able to identify this issue, reduce the dissipation,
and demonstrate that this problem does not affect the energy
balance of the results presented in this paper.

Subgrid-scale dynamical processes including turbulent mixing and
convection are parameterized
as in Forget et al. (1999).  In practice, the boundary layer
dynamics are accounted for by
Mellor and Yamada's (1982) unstationary 2.5-level closure scheme
plus a ``convective adjustment''  which rapidly mixes the atmosphere in
the case of unstable temperature profiles. Turbulence and convection mixes
energy (potential temperature), momentum (wind), and tracers (gases and aerosols).
The subgrid-scale orography and gravity wave drag schemes of the present-day
Mars GCM (Forget et al. 1999) were not applied.
\nocite{Forg:99,Mell:82,Forg:99}

Surface temperature evolution is governed by the balance between
radiative and sensible heat fluxes
(direct solar insolation, thermal radiation from the atmosphere
and the surface, and turbulent fluxes) and thermal conduction in
the soil. The parameterization of this last process was
based on an 18-layer soil model solving the heat diffusion equation using
finite differences. The depth of the layers were chosen to capture
diurnal thermal waves
as well as the deeper annual thermal wave. Vertically
homogeneous soil was assumed. For most simulations,
the thermal inertia was set to 250~J~s$^{-1/2}$~m$^{-2}$~K$^{-1}$
everywhere (a value slightly higher than the mean value on present-day Mars,
 to account for the higher gaseous pressure in the pore space, which increases the
soil conductivity).
For the ground albedo, as well as for the topography,
we chose to keep the same
distributions as observed on Mars today. Both fields
may have been quite different
three billion years ago, but we assume that the present-day
values can be representative
of the range and variability of the Noachian-Hesperian eras. In
Section~\ref{sc:tsurf}, the sensitivity of our results to surface
thermal inertia and albedo is discussed.

\subsection{Radiative transfer in a thick CO$_2$ atmosphere}

\label{sc:cia}

Our radiative scheme is based on the correlated-k model, with
the absorption data calculated directly from high resolution
spectra computed by a line-by-line model from the HITRAN 2008
database (Rothman et al., 2009).
These were then converted to correlated-k coeficients for use
in the radiative transfer calculations.
\nocite{Roth:09}

In practice, at a given pressure and
temperature,  correlated-k coefficients in the GCM are interpolated from
a matrix of coefficients stored in a $6 \times 9$ temperature
and log-pressure grid: $T = {100, 150, 200, 250, 300,  350}$~K,
$p = 10^{-1}$,  10$^{0}$,  10$^{1}$, ....,  10$^{7}$~Pa.
We used 32 spectral bands
in the thermal infrared and 36 at solar wavelengths. Sixteen points were
used for the g-space integration, where g is the cumulated distribution
function of the absorption data for each band.
Rayleigh scattering by  CO$_2$ molecules
was included using the method
described in Hansen and Travis (1974), and using
the Toon et al.  (1989) scheme to compute the radiative transfer.
\nocite{Hans:74,Toon:89}

As mentioned in Section~\ref{sc:previous}, a key improvement of our radiative
transfer model compared to previous models is the use of an improved
parameterization for the CO$_2$ collision-induced
absorption (CIA). It was specially developed for the
present study  using the results of Baranov
et al. (2004) and Gruszka and Borysow (1998).
The method is described and justified
in Wordsworth et al. (2010).
The sublorentzian
profiles of Perrin and Hartmann (1989) were used for
the CO$_2$ far line absorption.

\nocite{Bara:04,Grus:98,Word:10co2,Perr:89}

\subsection{CO$_2$ ice condensation and  clouds}

\label{sc:modelcloud}

The model includes a parameterization to account for
the possible condensation of
CO$_2$ when temperatures fall below the condensation temperature $T_c$.
To compute $T_c$ as a function of pressure $p$ (Pa), we used the
expression provided by Fanale et al. (1982):
\nocite{Fana:82}

\begin{equation}
 T_c = -3167.8/[\ln(0.01p)-23.23]
\end{equation}
for $p<518000$~Pa,

\begin{equation}
T_c= 684.2-92.3\ln(p)+4.32\ln(p)^2
\end{equation}
for $p>518000$~Pa
(518000~Pa is the pressure of the triple point of CO$_2$).

As on Mars today,
CO$_2$ can directly form on the ground. When the surface temperature
falls below the condensation temperature, CO$_2$ condenses, releasing
the latent heat required to keep the solid-gas interface at the
condensation temperature. Conversely, when CO$_2$ ice is heated,
it partially sublimates to keep its temperature at the frost point.
When CO$_2$ ice is present, the surface albedo and emissivity are
set to 0.5 and 0.85 respectively.
On present-day Mars, this albedo varies a lot in space and time.
 0.5 is consistent with the various
estimations of the average CO$_2$ ice cap albedo used to fit the seasonal pressure
variations induced by the condensation and sublimation in the polar caps (Forget
et al. 1998, Haberle et al. 2008)
\nocite{Forg:98,Habe:08pss}

In our simulations, CO$_2$ ice clouds also form in the atmosphere. Such clouds
have been observed on present-day Mars
during the polar night in the lower atmosphere (see
e.g. Pettengill and Ford, 2000, Tobie et al. 2003)
and in the equatorial mesosphere around 60-80~km (Montmessin et al. 2007b,
M\"a\"att\"anen et al. 2010).
These mesospheric  CO$_2$ ice clouds may share some
similarities with those predicted in our early
Mars simulations. However they form more rarely (probably
because the atmosphere is not
often below the frost point
(Gonzalez-Galindo 2010, Forget et al. 2009))  and at significantly
lower pressures (0.1 to 0.01~Pa). In spite of this, the observed optical depths
and particle radii are large (typically around 0.2 and 3~$\mu$m,
respectively), suggesting
that the growth of CO$_2$ ice particles in a CO$_2$~gas atmosphere is a very
efficient process.
\nocite{Pett:00,Tobi:03,Mont:07clouds,Maat:10,Gonz:10,Forg:09}

In theory, to properly model the formation of such clouds, one must take into
account various microphysical processes such as supersaturation,  nucleation and
crystal growth (Wood 1999, Colaprete and Toon 2003).
In most of our simulations, however, we  assumed
that condensation would occur as soon as the temperature $T^*$
predicted from the dynamical and radiative cooling rates dropped
below $T_c$. In section~\ref{sc:microphysic}, we also present results from a test case
in which a 30$\%$ supersaturation is required before condensing.
\nocite{Wood:99,Cola:03a}

At each timestep, the mass mixing ratio $m$ (kg/kg)
of condensed ice in a model box  (or its evolution $\delta m$
if ice is already present) is simply deduced from the amount of
latent heat needed to keep $T=T_c$:
\begin{equation} \delta m= \frac{c_p}{L} (T_c - T^*)\end{equation}

with  $\delta m$ the mass mixing ratio of ice
that has condensed or sublimated ($>$0 when condensing),
$c_p$  the specific heat at constant pressure (we took
735.9~J~kg$^{-1}$~K$^{-1}$) and $L$ the  latent
heat of CO$_2$ ($5.9~\times~10^5$~J~kg$^{-1}$).

CO$_2$ ice is transported by the large-scale circulation and by turbulent and
convective mixing. The transport scheme used in our GCM
is a  ``Van-Leer~I'' finite volume scheme
(Hourdin and Armengaud, 1999).
\nocite{Hour:99}

To estimate the size of the cloud particles,
we assumed that the number mixing
ratio of cloud condensation nuclei [CCN] (kg$^{-1}$) is constant throughout
the atmosphere. Assuming that the cloud particle size
distribution is monodisperse in each box, the cloud particle radius $r$
is then given by:

\begin{equation}
r=  (\frac{3 m}{4\pi \rho \ \mbox{[CCN]} })^{1/3}
\end{equation}

with $\rho$ the CO$_2$ ice density, set to 1620~kg~m$^{-3}$ in our
model.

Once $r$ is known, the cloud particle sedimentation velocity $w$
is calculated using Stokes law
(Rossow, 1978).
Sedimentation is then computed separately from the
transport, but  using a similar Van-Leer~I transport scheme in the vertical.
\nocite{Ross:78}

Sedimentation and condensation are strongly coupled. Within one
physical timestep (1/48 of a sol, or 1850~s in our model), we found that
a significant part of the ice that forms within a cloud
can sediment to the layers below, where it can condense at
a different rate or
even sublimate.  For these reasons, in the model,
condensation and sedimentation schemes are coupled together and
integrated with a sub-timestep equal to $1/20$ of the physical timestep.

The cloud particle sizes are also used to calculate the cloud
radiative effect.
Refractive indices for CO$_2$ ice are taken from
Hansen (2005) and Mie theory is assumed to retrieve the single scattering
properties. Scattering and absorption are computed at both solar and thermal
infrared wavelengths using the  Toon et al.  (1989) scheme.
\nocite{Hans:05,Toon:89}

[CCN] is clearly a key parameter which directly controls
the properties and the impact of the modeled
clouds.  Exploring the sensitivity of the results
to this poorly known parameter
allows us to account for most of the uncertainties related to the CO$_2$ ice
clouds microphysics and particles size distribution.
What is the possible range of [CCN]? On the Earth, the number
mixing ratio of cloud condensation nuclei in the troposphere ranges between
10$^{6}$~kg$^{-1}$ (for low saturation in clean polar air) and
10$^{10}$~kg$^{-1}$ (polluted air mass)(Hudson and Yun, 2002,Andreae, 2009). It
is significantly lower for icy cirrus clouds ($<$10$^{4}$~kg$^{-1}$)
(e.g. Demott et al. 2003) .
Even in the absence
of surface or chemical sources, a minimum number of nuclei would be
provided by meteoritic dust and smoke particles, which have
been suggested to be possible condensation nuclei
for terrestrial Polar Mesospheric Clouds (Gumbel and Megner, 2009). Such
particles must have been abundant four billion years ago.
In this paper, we explore [CCN] values ranging from 10$^{2}$~kg$^{-1}$
to 10$^{8}$~kg$^{-1}$, with a baseline value of  10$^{5}$~kg$^{-1}$.
\nocite{Huds:02,Gumb:09,Andr:09,Demo:03}

\subsection{Atmospheric dust}
\label{sc:dustmodel}

Early Mars winds were probably  able
to lift and transport mineral dust from the surface. If the planet was dry
enough, and in the absence of oceanic sinks as on the Earth, we must take into
account the possibility that the atmosphere was laden with
mineral aerosols, as on
Mars today.
To explore the impact of atmospheric dust on the early Mars climate, we added a
second type of aerosol in addition to the CO$_2$ ice clouds particles.
We assume that the dust is similar to that observed on present-day Mars and
use Wolff et al. (2009) optical properties, with a constant  effective
radius of 1.5~$\mu$m.
Instead of simulating the lifting and transport of dust by the winds, we
prescribe the dust mixing ratio in the atmosphere and assume that it does not
vary with time.  The column averaged dust optical depth at the mean
pressure level is horizontally uniform.  In the vertical,
the dust mixing ratio $q$
is taken to be constant in the lower
atmosphere up to a level above which it rapidly declines, as in Forget et al.
(1999):

\begin{eqnarray}
q = q_0 \exp\left\{ 0.007\left[1 - (p_0/p)^{(70\:{\rm km}/z_{\rm max})}
\right] \right\}
 &  \; \; \; p \leq p_0 \\
q = q_0  & \; \; \; p > p_0
\label{eq:newconrath}
\end{eqnarray}

where $q_0$ is a constant determined by the prescribed optical depth
at the global average surface pressure $p_0$, and $z_{\mbox{\scriptsize max}}$
the altitude (km) of the top of the dust layer. In this paper
simulations were performed with $z_{\mbox{\scriptsize max}}=30$ and $100$~km
(see Section~\ref{sc:dustresults}).

\section{Results}
\label{sc:results}

To explore the range of climates that may have occurred on a planet
like early Mars with a thick CO$_2$ atmosphere, we have performed multiple
simulations with different values of key model parameters~:
1) mean surface pressure, 2) cloud microphysics parameters (i.e. cloud
condensation nuclei density), 3) obliquity and eccentricity,
4) surface properties, 5) atmospheric dust loading. The impact of
water vapor is also discussed.
Solar luminosity is set to 75\% of the present value (see
Section~\ref{sc:faint} for a discussion of this assumption).
The solar spectrum was assumed to be the same as today.
All results correspond to the last year of a 10-year
simulation. In all cases, the modelled planet has reached equilibrium
by this time (results
are repeatable from year to year), except in the case of permanent CO$_2$
condensation, which induces a slow decrease of the atmospheric mass and pressure
(atmospheric collapse; see below).

\subsection{Surface temperatures and CO$_2$ ice caps.}

\label{sc:tsurf}

In this section, we describe simulations performed assuming a circular orbit,
25$^{\circ}$ obliquity as on present-day Mars, no dust,
and a constant cloud condensation nuclei number density
[CCN] set to $10^5$~kg$^{-1}$.

\begin{figure}
  \begin{center}
   \includegraphics[width=8.4cm,angle=-0,clip]{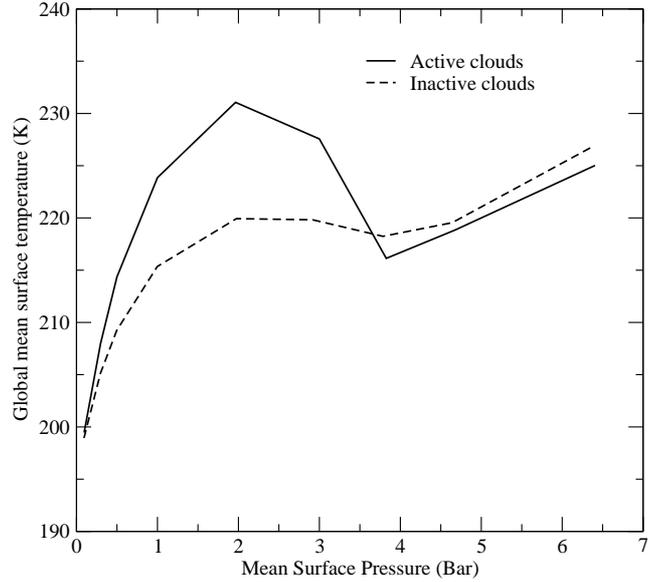}
    \caption{
\label{fg:ts_ps}
Global mean annual mean surface temperature (K) as a function of surface pressure in our baseline simulations (obliquity = 25$^{\circ}$, [CCN]=10$^5$~kg$^{-1}$, circular orbit) with and without radiatively active CO$_2$ ice clouds. }
  \end{center}
\end{figure}
\begin{figure}
  \begin{center}
   \includegraphics[width=8.4cm,angle=-0,clip]{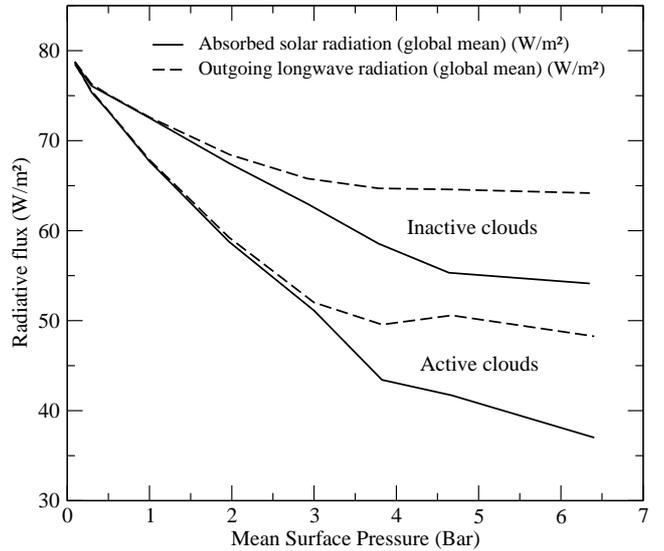}
    \caption{
\label{fg:olr}
Global mean annual mean radiative budget for the same simulations as in
Figure~\ref{fg:ts_ps} with or without
radiatively active clouds. After 10 years,
simulations with surface pressure higher than 3 bars are out of
equilibrium because of atmospheric collapse and the constant
release of latent heat at the surface.}
  \end{center}
\end{figure}

Figure~\ref{fg:ts_ps} presents global/annual mean surface
temperature (K) as a function of mean surface pressure $Ps$.
Results obtained without taking into account the radiative effects of CO$_2$
ice clouds (the clouds are assumed to be ``transparent'' at all wavelengths)
are in agreement with previous 1D model calculations
performed with the same radiative transfer parameterization (Wordsworth et al.
2010). Surface temperature increases up to $Ps=2$~bar. Above 2-3 bar
Rayleigh scattering by CO$_2$ gas more than compensates for the
increased thermal infrared opacity of the atmosphere. Increasing the
atmospheric thickness does not result in an increase of the mean surface
temperature.  Taking into account the radiative effect of CO$_2$ ice clouds
results in a global warming of the surface by more than 10~K resulting from the
CO$_2$ ice cloud scattering greenhouse effect (Forget and Pierrehumbert, 1997).
In the GCM, this effect is significant, but not as much as it could have
potentially been according to the estimation by Forget and Pierrehumbert (1997).
This is
mostly because the cloud opacities remain relatively low compared to what was
assumed in that study (see below). Figure~\ref{fg:olr} shows the radiative
budget corresponding to Figure~\ref{fg:ts_ps} simulations.
One can see that the clouds strongly raise the planetary albedo
and thus decrease the absorbed solar radiation.
On this plot, one can verify that the absorbed solar energy is
equal to the emitted infrared energy (i.e. the simulated planet is
in radiative equilibrium as expected), except at high pressure
when the atmosphere collapses on the surface and releases
latent heat, as detailed below.

\nocite{Word:10co2,Forg:97}

\begin{figure*}
  \begin{center}
   \includegraphics[width=17.cm,angle=-0,clip]{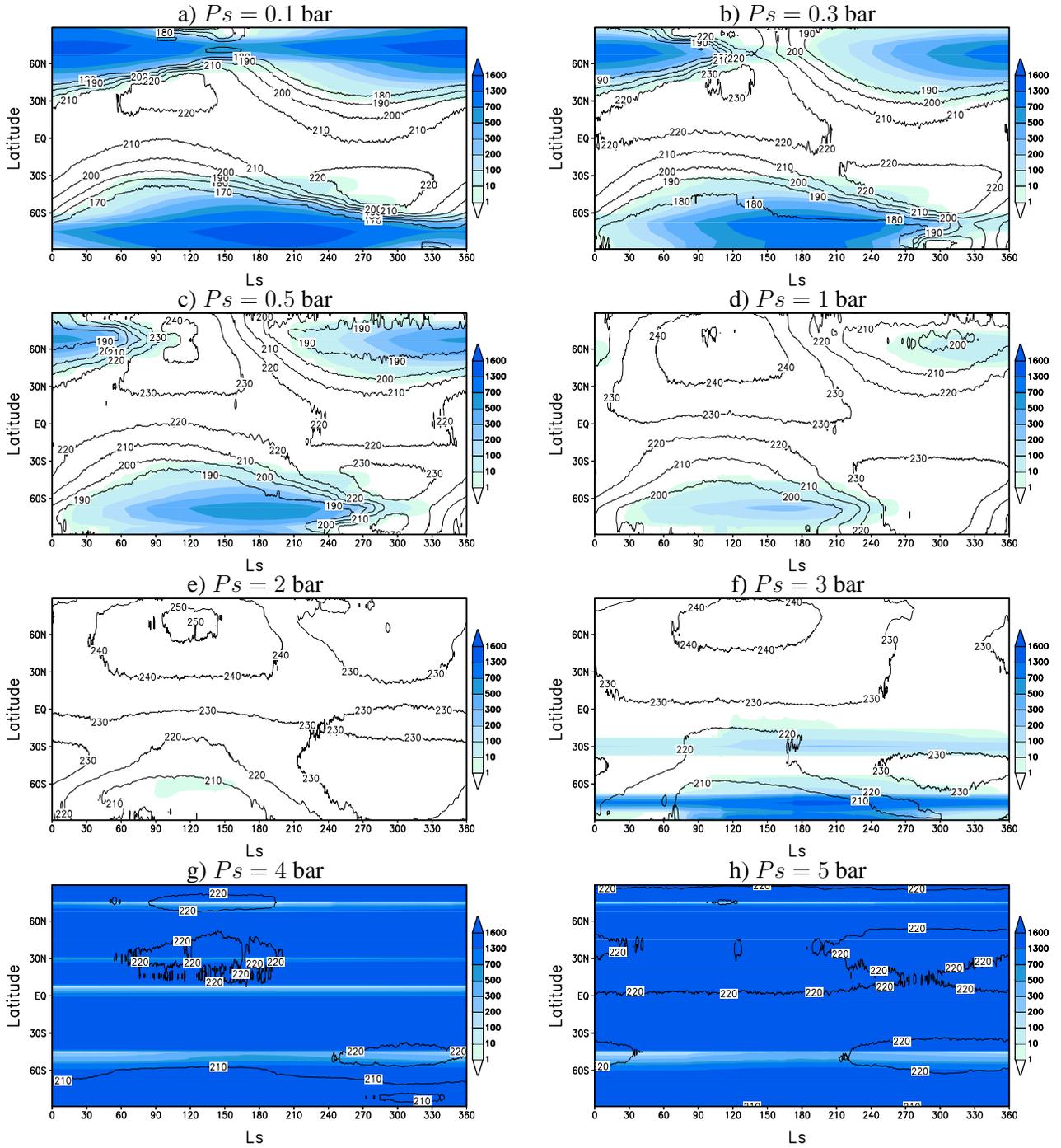}
    \caption{
\label{fg:ts_ls}
Zonal mean surface temperatures (contours, K) and surface CO$_2$ ice
(kg~m$^{-2}$) as a function of solar longitude $L_s$ (degree) for various
mean surface pressures $Ps$
(obliquity=25$^{\circ}$, [CCN]=10$^5$~kg$^{-1}$, circular orbit).
    }%
  \end{center}
\end{figure*}

Figure~\ref{fg:ts_ls} presents the seasonal and latitudinal variation of
zonal mean surface temperature and surface CO$_2$ ice for the different
surface pressure experiments. The maximum accumulation of CO$_2$ ice is not
found exactly at the poles because CO$_2$ ice clouds
are predicted to be especially thick there, and their effect on the incident
thermal infrared flux tend to limit the surface cooling and thus the
surface condensation.

With $0.5 \leq Ps \leq 2$~bar, seasonal CO$_2$ ice caps are
predicted to form at high
latitudes during fall, winter and spring as in the Mars northern hemisphere
today.  For $Ps \leq 0.3$~bar, however, permanent surface
CO$_2$ ice glaciers are predicted to form at high latitudes
in both the radiatively active and
inactive cloud cases.
In these simulations, the atmosphere is collapsing
and the results after 10 years may not represent a realistic long-term solution.
Many more years would be required to reach a steady state in which
the permanent CO$_2$ ice caps would be in solid-gas equilibrium with
the atmosphere, with a significant part of the atmosphere trapped on
the surface.

\begin{figure*}
  \begin{center}
   \includegraphics[width=17.cm,angle=-0,clip]{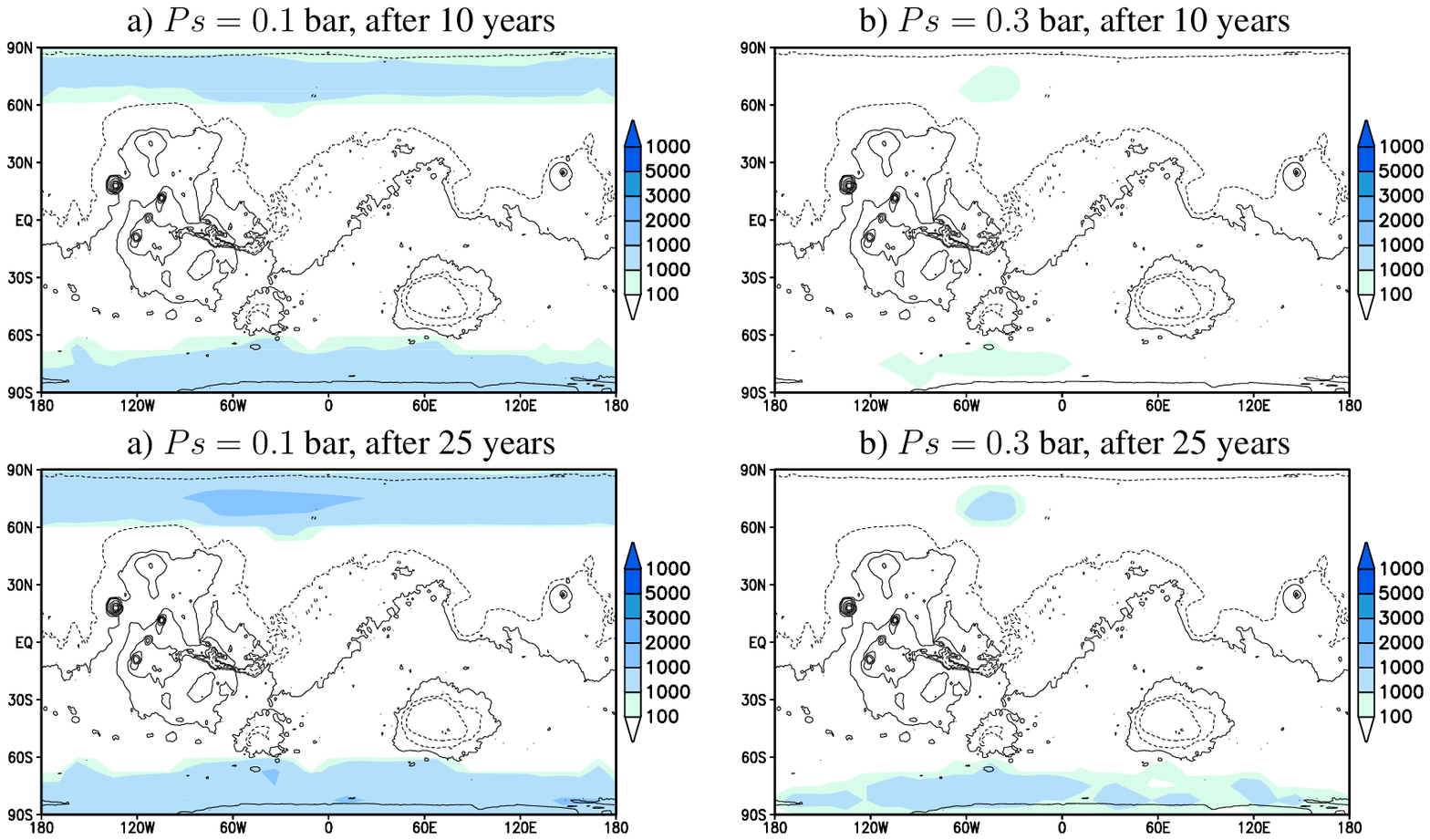}
    \caption{
\label{fg:co2min1}
Maps of the permanent surface CO$_2$ ice deposits (kg m$^{-2}$)
for the $Ps=0.1$ and
$Ps=0.3$~bar simulations, also shown in Figure~\ref{fg:ts_ls}.
In practice, these maps show the yearly minimum CO$_2$ ice mass recorded during
the 10th and 25th simulated years, to show where the ice never sublimates during the
year. Contours illustrate topography (contours below zero elevation are dotted)}
  \end{center}
\end{figure*}

Figure~\ref{fg:co2min1} show maps of
the extension of the permanent CO$_2$ caps after 10 and 25~years.
Between 1 and 25~years, the atmospheric mass collapses
by 1\% per year in the (initially) 0.1-bar
simulations, whereas the pressure drops by only 0.032\% per year in the 0.3-bar
case.
To properly simulate an equilibrated atmosphere/permanent CO$_2$ ice cap
system, it would also be necessary to take into account the ability of CO$_2$
ice glaciers to flow and spread (CO$_2$ ice  is known to
be much softer than H$_2$O ice at the same temperature (Kreslavsky and Head,
2011, Durham et al. 1999)), and probably to include the effect of slopes and
roughness (Kreslavsky and Head, 2005).

\nocite{Kres:11,Durh:99,Kres:05}

The formation of permanent polar caps at low pressure is also very dependent
on obliquity. This is discussed in detail in section~\ref{sc:obliquity}.

\begin{figure*}
  \begin{center}
   \includegraphics[width=17.cm,angle=-0,clip]{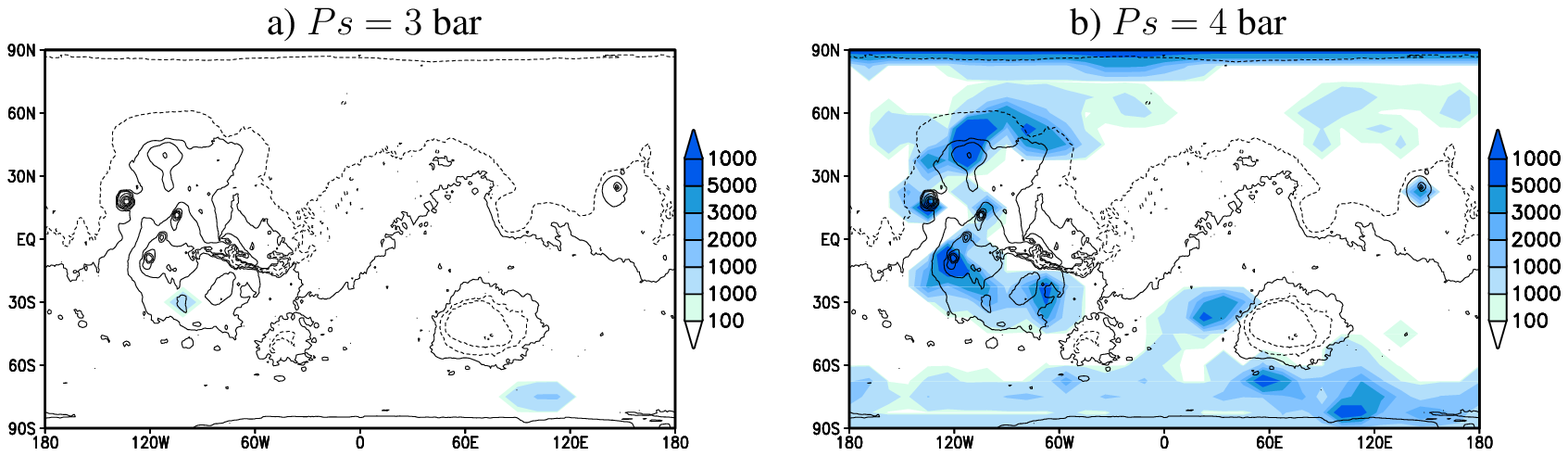}
    \caption{
\label{fg:co2min2}
Same as Figure~\ref{fg:co2min1}, but for the $Ps=$3 and
4~bar simulations (also shown in Figure~\ref{fg:ts_ls}).
These show the yearly minimum  CO$_2$ ice mass  (kg m$^{-2}$) recorded during
the 10th year of the simulation.}
  \end{center}
\end{figure*}

For $Ps \geq 3$~bar, permanent surface
CO$_2$ ice glaciers are also predicted to form on the surface
in both the radiatively active and inactive cloud cases.
CO$_2$ ice tends to condense permanently
in the colder areas at high altitudes and latitude.
In the $Ps=3$~bar case, the permanent CO$_2$ ice glaciers are restricted to a
couple of locations on the Tharsis bulge near 30$^{\circ}$S-100$^{\circ}$W,
and in the southern high latitudes around 75$^{\circ}$S-100$^{\circ}$E
(Figure~\ref{fg:co2min2}a). For $Ps=4$~bar and more, the permanent CO$_2$ ice
sheets are much more extensive, covering high latitudes and most of the Tharsis bulge
(Figure~\ref{fg:co2min2}b).

The fact that permanent CO$_2$ ice cap form at either low or high pressure can
be interpreted as follows~: At low pressures, the greenhouse effect and heat
transport are weak and thus temperatures easily reach the frost point. At high
pressures the increase in the frost point temperature overcomes the increased
greenhouse effect and CO$_2$ also condense easily. In addition,
heat transport is then so efficient that horizontal temperature
gradients are small.  Consequently, CO$_2$ ice can form in non-polar regions.

The formation of extensive permanent and seasonal polar caps at high pressures
buffers the surface temperature and explains the increase of global mean
surface temperature for $Ps \geq$4~bar in Figure~\ref{fg:ts_ps}. In these
simulations, thick CO$_2$ ice clouds form just above the surface.
Such clouds tend to cool the surface (Mischna et al., 2000) because they reflect
solar radiation throughout their entire depth
(the planetary albedo is increased) whereas only the upper layers can
contribute to reflect the thermal infrared flux
emitted by the atmosphere below them.
As a result, the
radiatively active clouds simulations show mean surface temperatures even
lower than when the clouds are assumed to be transparent.
\nocite{Misc:00}

\begin{figure*}
  \begin{center}
   \includegraphics[width=17.cm,angle=-0,clip]{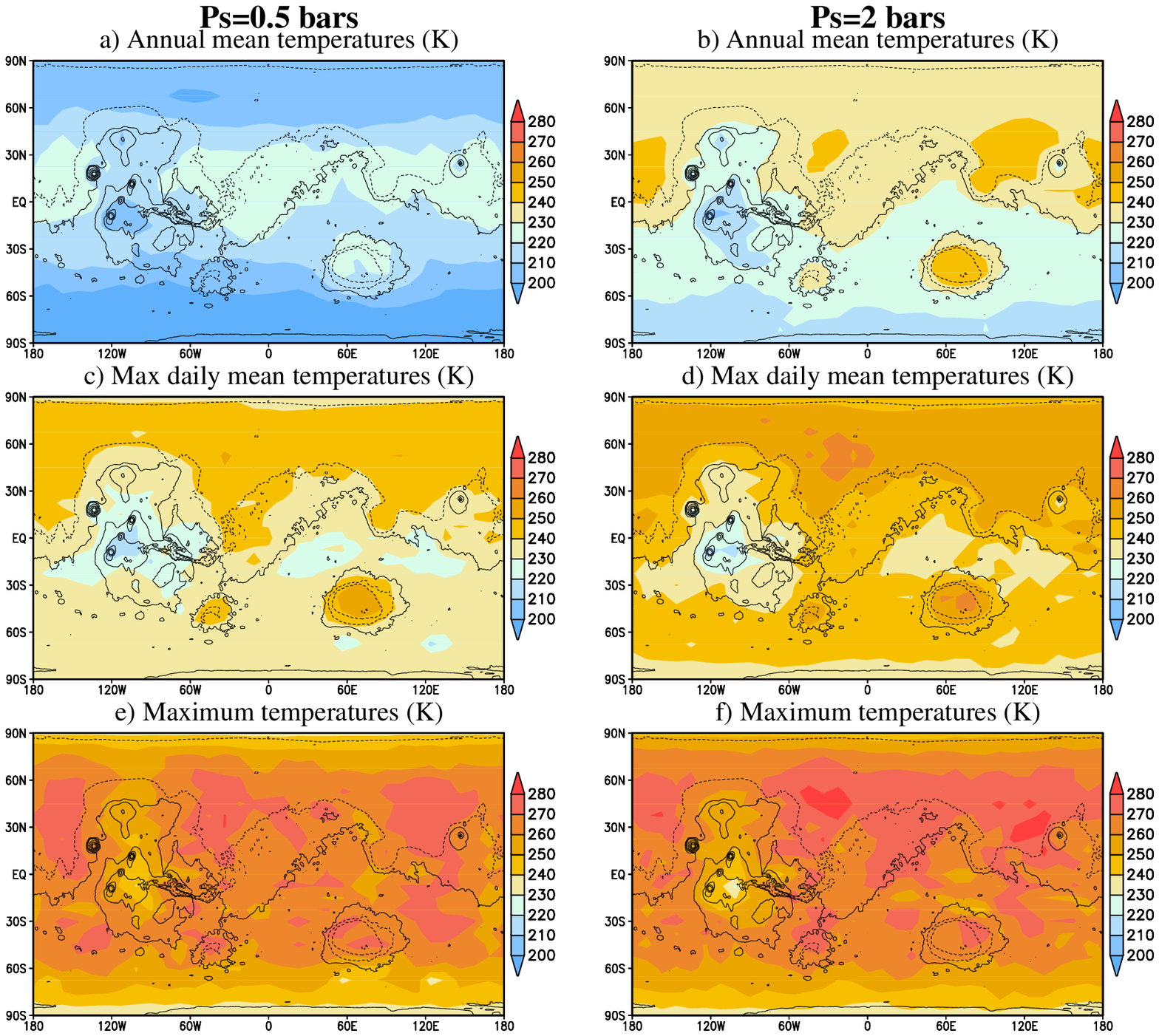}
    \caption{
\label{fg:mapts}
Surface temperatures (K) in our baseline simulations
(obliquity=25$^{\circ}$, [CCN]=10$^5$~kg$^{-1}$, circular orbit)
for mean surface pressure 0.5~bar and 2~bar. Ground albedo
is as on present day Mars. Thermal inertia is set to 250~J~s$^{-1/2}$~m$^{-2}$~K$^{-1}$ everywhere.
    }%
  \end{center}
\end{figure*}

\begin{figure}
  \begin{center}
   \includegraphics[width=8.4cm,angle=-0,clip]{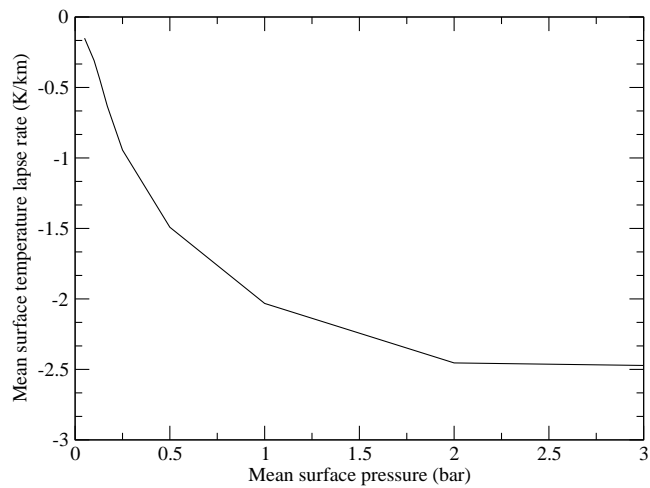}
    \caption{
\label{fg:lapserate}
Mean surface temperature
variation with topography (K/km) as a function of mean
surface pressure (bar). This lapse rate is obtained by performing a linear
regression between the annual mean surface temperature obtained below
30$^{\circ}$ latitude (to avoid the influence
of the polar caps) and the local topography in our baseline cases
(obliquity = 25$^{\circ}$, [CCN]=10$^5$~kg$^{-1}$, circular orbit).
 }
  \end{center}
\end{figure}

Figures~\ref{fg:mapts}a and \ref{fg:mapts}b present annual mean surface
temperature maps in the  more realistic $Ps=$0.5 and 2~bar cases.
One can notice
that surface temperatures strongly depend on the local topography,
especially in
the 2~bar
 simulations in which the lowest plains are the warmest places on Mars.
This is not due to a stronger greenhouse warming where the atmosphere is
thicker. As illustrated on Figure~\ref{fg:ts_ps}, the
increase of surface temperature with CO$_2$ pressure due to the
greenhouse effect is weaker at high pressure and negligible above 2 bars.
The opposite is found in our maps.

In fact, Earthlings are familiar with this situation, which
results from adiabatic cooling and warming of the atmosphere when it moves
vertically and the fact that the atmosphere can influence the surface
temperature when it is dense enough.
On present-day Mars, the atmosphere is too
thin to affect the surface temperature. The
local topography has no significant effect on surface temperature.
At which atmospheric pressure does the transition between a ``present-day Mars
regime'' and an ``Earth-like regime'' occur? To address this question, we have
computed the mean surface temperature lapse rate as a function of mean surface
pressure (Figure~\ref{fg:lapserate}). The lapse rate is simply estimated by performing a
linear regression between annual mean surface temperatures and ground altitude
at every grid point for latitudes lower than 30$^{\circ}$, to avoid the
influence of seasonal CO$_2$ ice condensation. We can see that the surface
temperature lapse rate quickly decreases with increasing pressure.
The effect is already
significant  (-1.5~K/km) for $Ps=0.5$ bar, and reaches an asymptotic value
(-2.5~K/km) above about 2~bars. This is half the theoretical dry
atmospheric lapse rate equal to -5.05~K/km in our model.
This can be compared to the Earth case, where surface temperature lapse
rates ranging between around -3 and -6.5~K/km have been reported (See Minder et
al., 2010, and references therein), while the reference dry atmospheric
lapse rate is -9.8~K/km. However, on the Earth water condensation can decrease the lapse
rate by a factor of two.
\nocite{Mind:10}

\begin{figure*}
  \begin{center}
   \includegraphics[width=17.cm,angle=-0,clip]{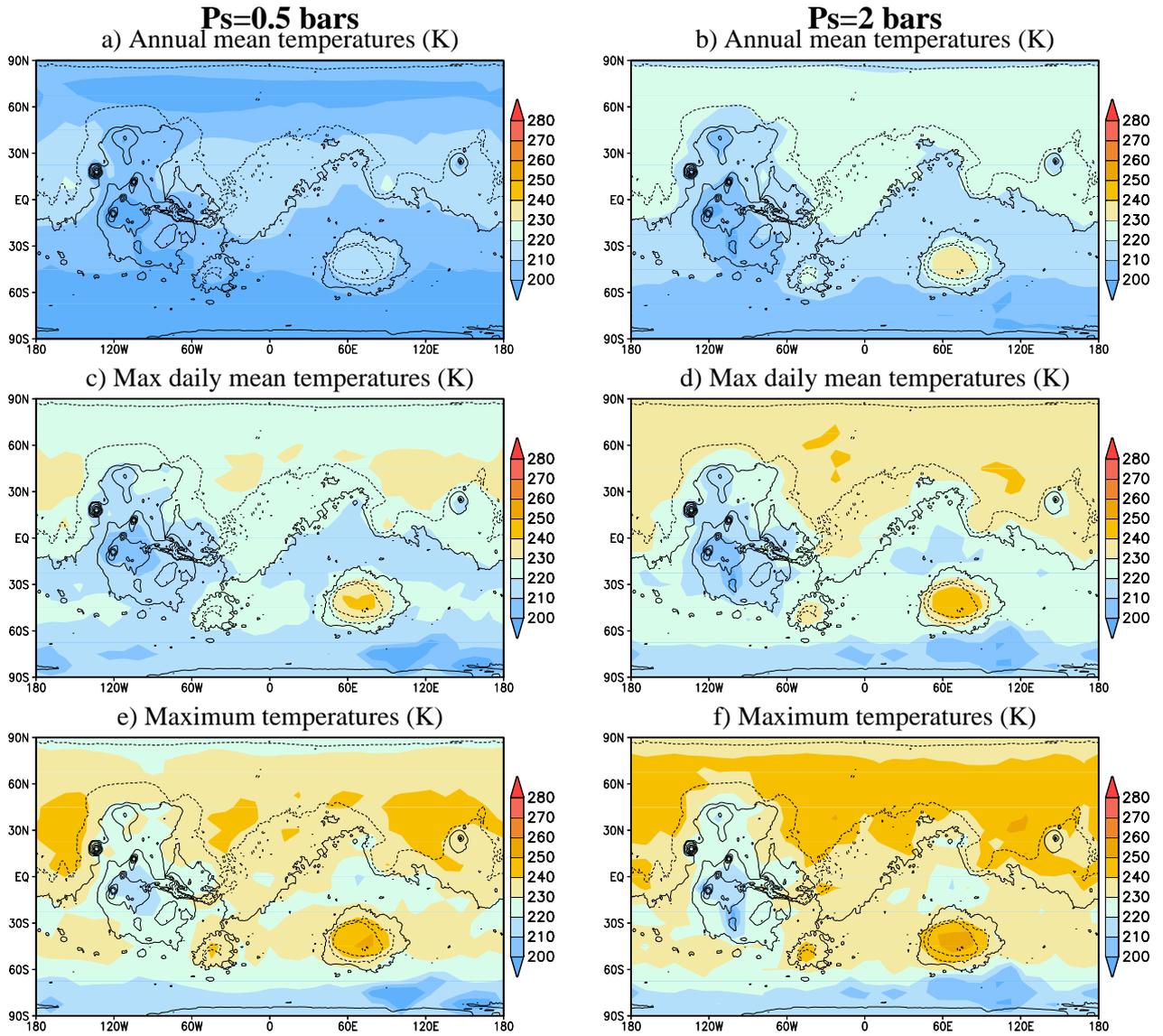}
    \caption{
\label{fg:maptsice}
Same as Figure~\ref{fg:mapts}, but for an ``ice-covered'' Mars, in which the
ground albedo and thermal inertia are set to 0.4 and
1000~J~s$^{-1/2}$~m$^{-2}$~K$^{-1}$ everywhere.
    }%
  \end{center}
\end{figure*}

In our simulations, annual mean surface temperatures are always significantly
below 0$^{\circ}$C.
In theory, this means that potential oceans and near-surface
groundwater should be permanently frozen in ice sheets and permafrost.
Could surface lakes and rivers form?
To investigate this possibility, a better criterion
is the average temperature in summer.
On the Earth, for instance, intense
fluvial activity takes place in Northern Siberia in the summertime,
despite annual mean surface temperatures below -15$^{\circ}$C (258~K).
Maps of peak summertime daily mean temperatures are shown in
Figures~\ref{fg:mapts}c and \ref{fg:mapts}d.
It appears
that, unlike those in Siberia, these temperatures never reach the
melting point of pure water in our baseline simulations.
In fact, the climate predicted by our model should be, at best, analogous to
Antarctica's upper Dry Valleys (Marchant and Head, 2007).
Water may melt, but only for a few hours
during a few afternoons in summer (as on Mars today).
The locations where above-freezing temperatures are predicted are
shown in Figures~\ref{fg:mapts}e and \ref{fg:mapts}f, which
presents the maximum surface temperatures
in the $Ps=$0.5 and 2~bar cases. Melting would be possible in many areas
below 60$^{\circ}$ latitude.
Although the mean temperatures are lower in the 0.5~bar than
in the 2~bar simulation, the diurnal amplitude is larger and the maximum
temperatures are not very dependent on pressure.
The amount of water which can melt during such short episodes
should be limited (see Wordsworth et al., 2012).
One can speculate that the meltwater could create supraglacial channels
below the glaciers, maintain perennially ice-covered lakes
(notably by bringing energy in the form of the
latent heat of fusion; McKay et al. 1985), or that brines
may play a major role.
\nocite{McKa:85,Marc:07}
However, one must keep in mind that in our baseline simulations we modeled a
dry, desertic planet Mars. First, we do not take into account the latent heat
losses which tend to cool the surface if ice were present. Second,
surface ice could possibly have higher albedo and
thermal inertia than bare ground. To assess the sensitivity of our results to
surface properties and the presence of ice, we performed additional
simulations with a global albedo of 0.4 (instead of the present day Mars ground
albedo which is near 0.22 on average)
 and a global thermal inertia of
1000~J~s$^{-1/2}$~m$^{-2}$~K$^{-1}$ (instead of
250~J~s$^{-1/2}$~m$^{-2}$~K$^{-1}$ assumed for dry ground).
As shown in Figure~\ref{fg:maptsice}, annual mean temperatures
are lowered by about 8 and 12~K at 0.5 and 2~bar respectively. This
reflects the effect of lowering the albedo, since
thermal inertia has little effect on annual mean temperatures.
However, increasing the
thermal inertia dampens the maximum temperatures, which are lowered by
about 30~K. This suggests that extensive ice deposits would not reach
0$^{\circ}$C, and that,  as in Antarctica upper Dry Valleys today,
only ice lying on dark rocks may sometimes melt (Marchant and Head, 2007).
\nocite{Word:12,Marc:07}

\subsection{Atmospheric temperatures and CO$_2$ ice clouds}
\label{sc:cloud}

\begin{figure*}
  \begin{center}
   \includegraphics[width=17.cm,angle=-0,clip]{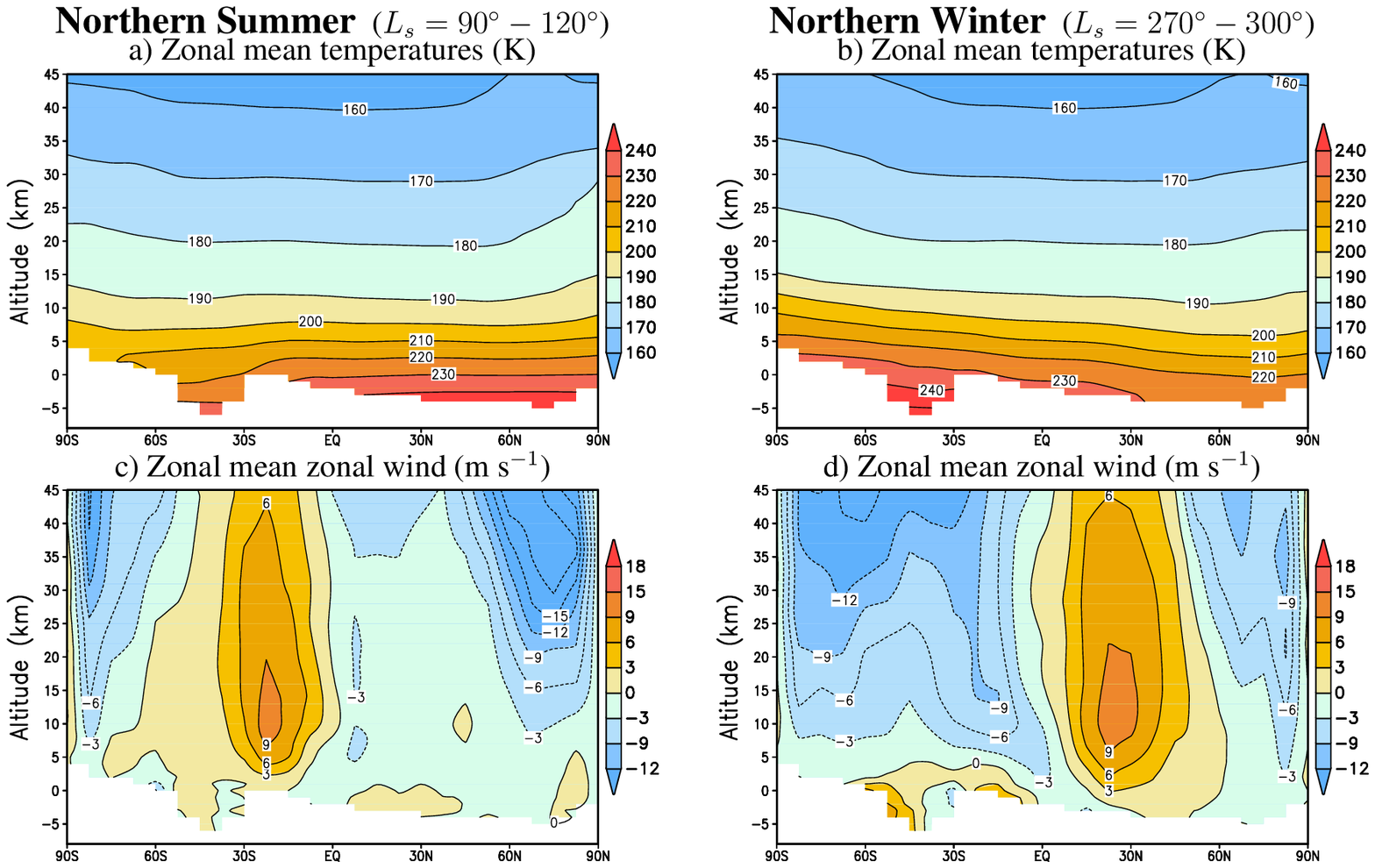}
    \caption{
\label{fg:sectionT}
Time-mean section of zonal-mean temperature (K) and zonal wind (m~s$^{-1}$) for
two opposite seasons in our baseline simulation with mean surface pressure
2~bar (obliquity=25$^{\circ}$, [CCN]=10$^5$~kg$^{-1}$, circular orbit).
The white areas represent grid points below the surface.  }%
  \end{center}
\end{figure*}

Figure~\ref{fg:sectionT} presents cross-sections of zonal mean
temperatures and zonal winds in northern summer
and winter in the 2~bar baseline simulation. At this pressure,
the atmospheric thermal structure
is relatively homogeneous, with little latitudinal and seasonal variations
compared to present-day Mars or even the Earth. Temperatures decrease
monotonically with altitude, even near the surface, as discussed above.
The small latitudinal gradients nevertheless induce a meridional circulation.
Plotting the mass streamfunction (not shown) reveals that it is
characterized by Earth-like overturning Hadley cells between 30$^{\circ}$S and
30$^{\circ}$N, and which extend up to about 15~km. In a given season,
the cross-equatorial cell with rising motion in the spring-summer hemisphere and
descending motion in the fall-winter hemisphere dominates.  The corresponding
zonal wind structure is characterized by an Earth-like subtropical prograde
jet in the winter hemisphere. The summer hemisphere retrograde jet is
much more Mars-like. In the northern hemisphere, the large longitudinal
variations in topography induce a strong stationary wave which modulates the
zonal wind.

\begin{figure*}
  \begin{center}
   \includegraphics[width=17.cm,angle=-0,clip]{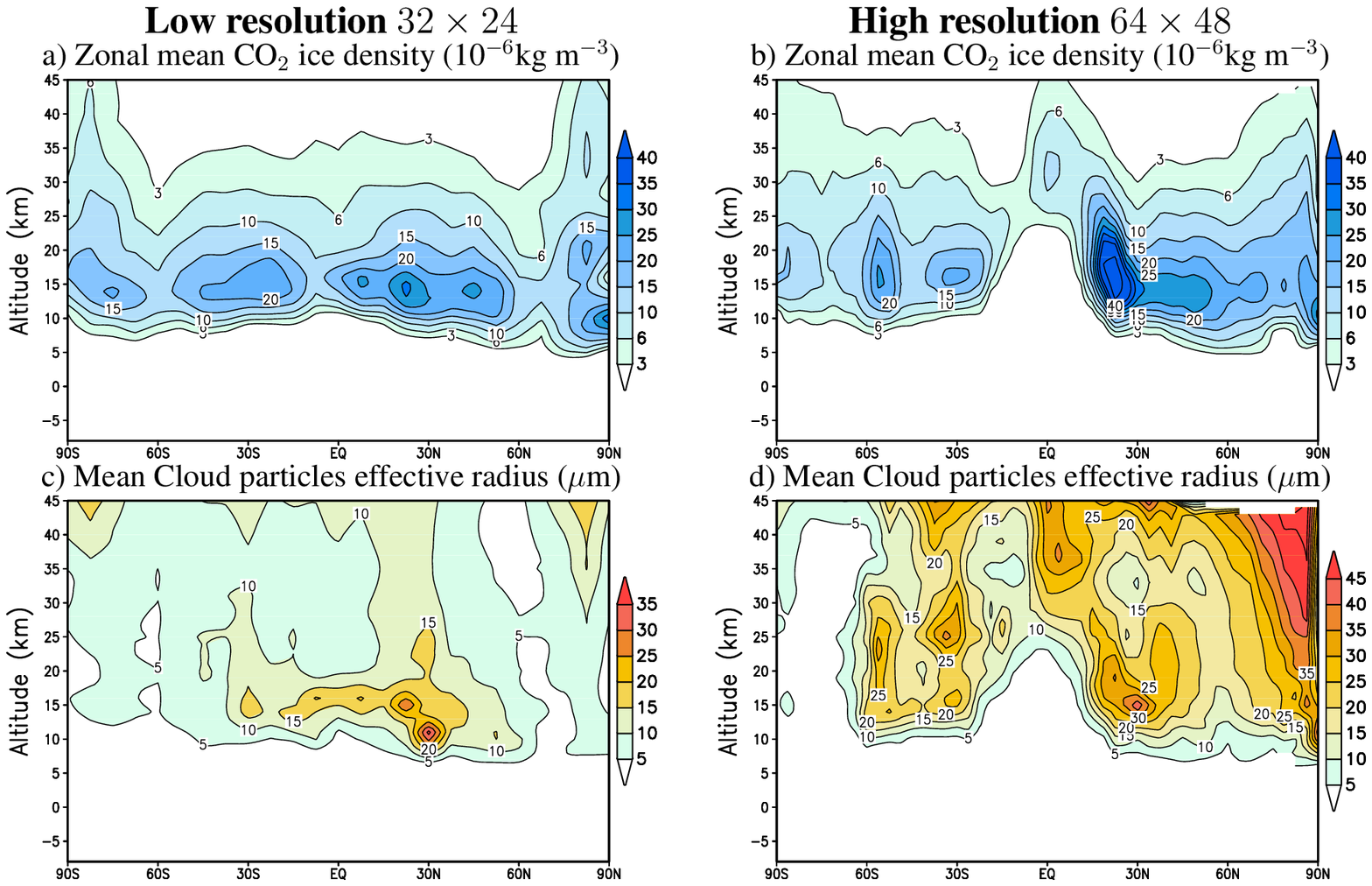}
    \caption{
\label{fg:section_clouds}
Annual-mean section of zonal-mean CO$_2$ ice cloud density
and particle radius,
for two simulations with different horizontal resolution
(mean surface pressure 2~bar, obliquity=25$^{\circ}$, [CCN]=10$^5$~kg$^{-1}$,
circular orbit). Particle radii are averaged with a relative weight proportional
to the CO$_2$ ice mixing ratio.
    }%
  \end{center}
\end{figure*}
\begin{figure*}
  \begin{center}
   \includegraphics[width=17.cm,angle=-0,clip]{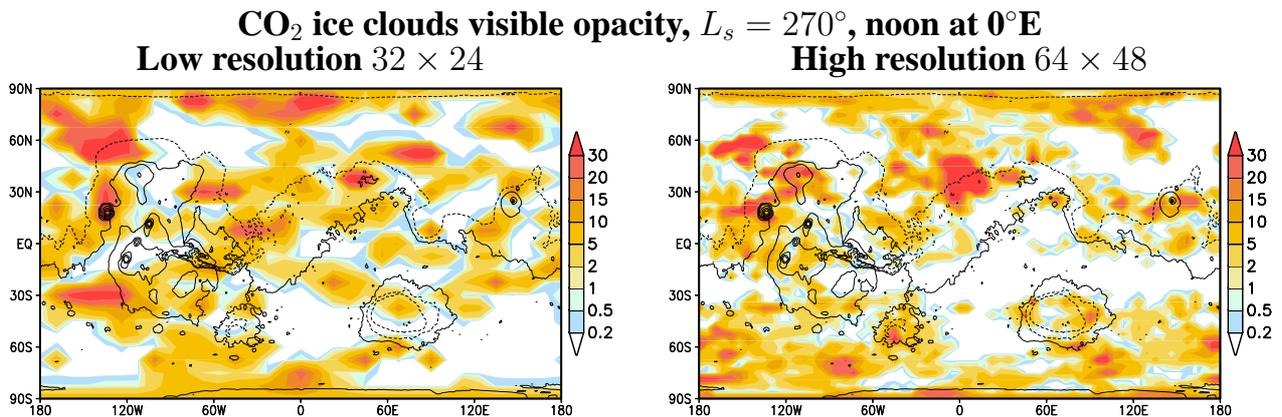}
    \caption{
\label{fg:map_cloud_inst}
An example of the instantaneous CO$_2$ ice clouds coverage for two simulations with
different horizontal resolution
(mean surface pressure 2~bar, obliquity=25$^{\circ}$, [CCN]=10$^5$~kg$^{-1}$,
circular orbit)
    }
  \end{center}
\end{figure*}
\begin{figure*}
  \begin{center}
   \includegraphics[width=17.cm,angle=-0,clip]{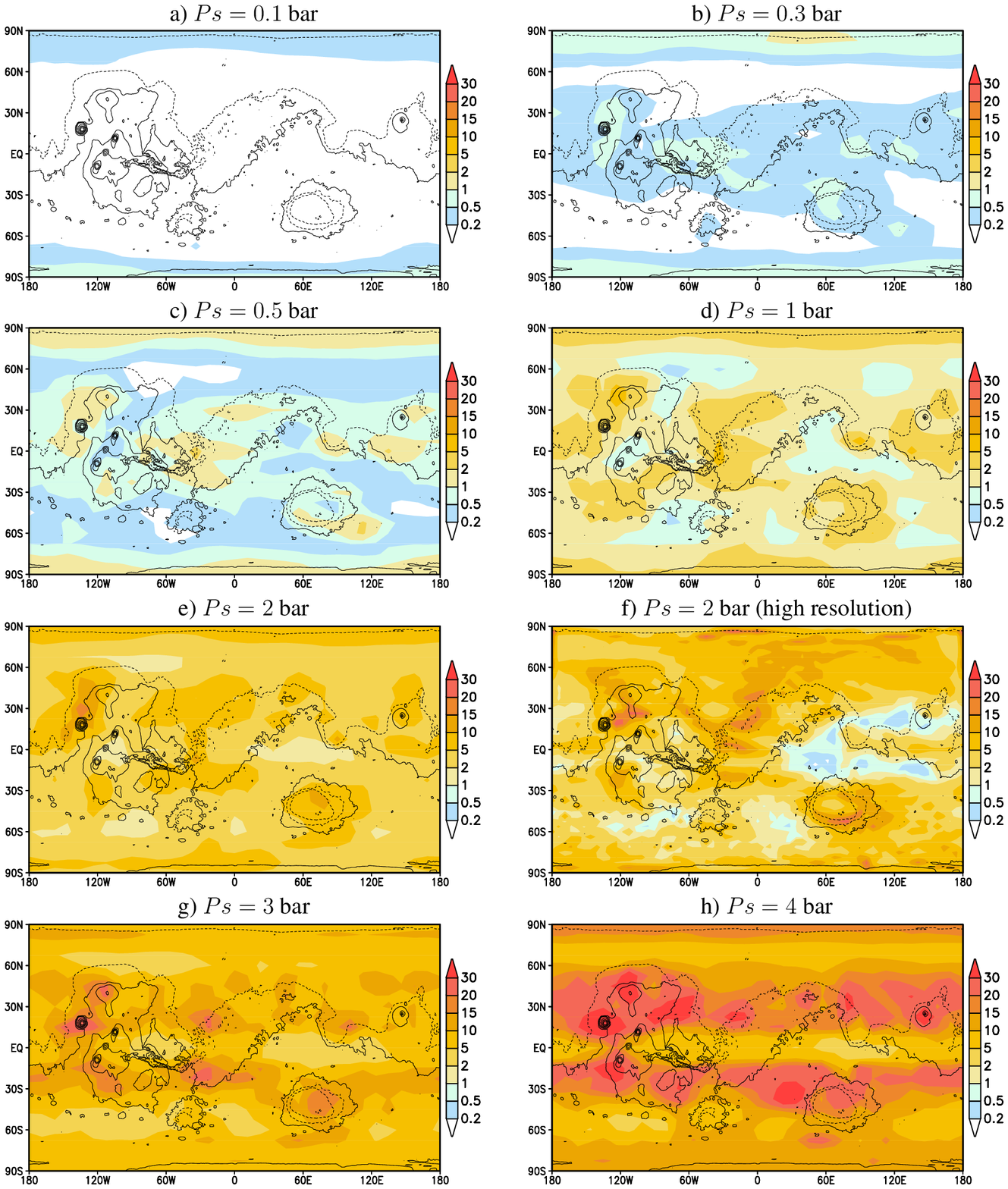}
    \caption{
\label{fg:co2cloud}
Map of annual mean CO$_2$ ice cloud optical depth
for various mean surface pressure $Ps$
(obliquity=25$^{\circ}$, [CCN]=10$^5$~kg$^{-1}$, circular orbit).
    }%
  \end{center}
\end{figure*}

In this 2-bar simulation, temperatures below the CO$_2$ condensation point are
predicted above about 11~km, and
CO$_2$ ice clouds form at all
seasons and latitudes
(Figure~\ref{fg:section_clouds}).  However, at any given time, CO$_2$
ice clouds typically cover about half of the planet (when counting
visible opacity above 0.2), as illustrated  in Figure~\ref{fg:map_cloud_inst}.
The locations of the clouds evolve constantly, with a combination of
transient cloud condensed and transported by travelling waves,
and stationary clouds forming above topography features
(such as Tharsis, the edge of Hellas or Arabia Terra). This
most likely results from resolved gravity waves. Is topography a key driver
of clouds and climate? To address this question, we performed a similar
simulation, but with topography removed. We found that the cloud coverage
is relatively similar,  with a global mean visible
opacity of 6 compared to 4.5 with topography. The mean surface temperature
is almost identical (231.5~K compared to 231.0~K).

Figures~\ref{fg:section_clouds}a and c show the annual
mean section of the cloud density and cloud particles radii in our
2-bar baseline simulation with topography.
The average particle radius is well above 10~$\mu$m, the
minimum size to readily scatter thermal infrared radiation and
induce the scattering
greenhouse effect (Forget and Pierrehumbert, 1997).
However, with a mean optical depth near 4.5 and fractional cloud cover
most of the
time, the clouds only induce a 10~K greenhouse warming, as
discussed above.
\nocite{Forg:97}

Because modeled clouds properties can be expected to be sensitive to model
resolution, we performed an additional 2-bar simulation
with doubled resolution in
latitude and longitude.  Results are compared in
Figures~\ref{fg:section_clouds}, \ref{fg:map_cloud_inst},
and~\ref{fg:co2cloud}e and f.
In the high-resolution simulation, the impact of
topography and gravity waves seems to be stronger. The
 cloud distribution (Figure~\ref{fg:co2cloud}f)
exhibits more structure in relation
to local topography.
Nevertheless, the mean cloud cover, optical depth, and particle
sizes are about the same on average
(average visible opacity of 4.6 vs 4.5 in the
baseline simulation). The mean surface temperature is only 1~K warmer, which is
not significant. Similarly, we explored the sensitivity to the model
vertical resolution by doubling the number of layers and thus the resolution
above the boundary layer (reaching a vertical resolution of about 2.5~km
between 5 and 25~km), and found very similar results.

Figure~\ref{fg:co2cloud} presents the annual mean
coverage in optical depth at other pressures in the baseline cases.
The average cloud visible optical depth $\tau$ growths
with mean pressure $Ps$,
but at the same time the
altitude of the bottom of the cloud layer $z$ decreases. For instance,
 $\tau$=0.7 and $z\simeq$14~km for $Ps$=0.5~bar;
$\tau$=1.8 and $z\simeq$11~km   for $Ps$=1~bar;
$\tau$=4.5 and $z\simeq$8~km  for $Ps$=2~bar;
$\tau$=9 and $z\simeq$5~km   for $Ps$=3~bar; and
$\tau$=16 and $z\simeq$0~km  for $Ps$=4~bar.
As mentioned above, the scattering greenhouse effect for low lying clouds is
reduced. Above 4 bars, the clouds cool the planet rather than warming it.

\subsection{Sensitivity to CO$_2$ ice cloud microphysics}
\label{sc:microphysic}

\subsubsection{Number of particles and particle sizes.}

\begin{figure}
  \begin{center}
   \includegraphics[width=8.4cm,angle=-0,clip]{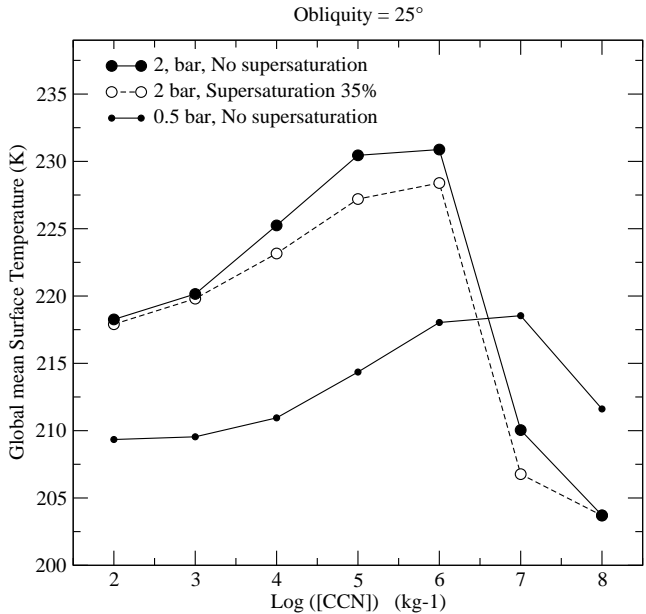}
    \caption{
\label{fg:nmix}
Global annual averaged
surface temperature (K) as a function of cloud condensation
nuclei number mixing ratio (kg$^{-1}$) in the 2~bar mean surface pressure cases
(obliquity = 25$^{\circ}$, circular orbit), and assuming either
no supersaturation or a 35~\% supersaturation to form CO$_2$ ice clouds (see
text).}
  \end{center}
\end{figure}

As explained in Section~\ref{sc:modelcloud}, we consider that
many of the uncertainties related to the
CO$_2$ ice clouds microphysics and
particle sizes
can be accounted for by varying the
prescribed  number mixing ratio of cloud condensation
nuclei [CCN]. We explored the sensitivity of the model to this
parameter in the $Ps=$0.5 and
2~bar cases by performing 7 simulations with [CCN] set to
10$^{2}$, 10$^{3}$,..., 10$^{7}$, 10$^{8}$~kg$^{-1}$ (see
Section~\ref{sc:modelcloud}). The primary effect of
decreasing [CCN] is to increase the
size of the cloud particles when they form.
On the one hand, this enhances the
sedimentation rate and thus decreases the mass of the cloud
(and thus its opacity). On the other hand, for a given cloud mass,
this affects the cloud opacity by changing the number of particles and
their single scattering properties. We found that the first effect
dominates: almost no clouds are present with [CCN]=10$^{2}$~kg$^{-1}$. With
higher [CCN], the average
cloud mass increased by a factor of about 3 for each order of
magnitude increase
in [CCN]. Meanwhile, the mean cloud particle sizes (computed by taking
into account sedimentation and sublimation below the condensation level) remains
near 10 to 20~$\mu$m. As a result, with  $Ps=$2~bar,
the mean cloud optical depth reaches 16, 67 and 156 for [CCN]=10$^{6}$,
10$^{7}$ and 10$^{8}$~kg$^{-1}$, respectively. In the last two cases, the cloud
optical depth is too thick to allow the scattering greenhouse effect to take
place (Forget and
Pierrehumbert, 1997). 
The clouds are so thick in the visible that albedo increases
outweigh the scattering of upwelling IR.
The clouds reflect most of the incoming solar radiation,
and their net effect is to cool the surface as seen in Figure~\ref{fg:nmix}.
With  $Ps=$0.5~bar, the mass of CO$_2$ ice clouds
is lower and the cloud warming is
maximized with more nuclei, up to 10$^{7}$~kg$^{-1}$ (Figure~\ref{fg:nmix})

\subsubsection{Supersaturation.}

As mentioned in Section~\ref{sc:previous},
based on the measured constraints on the critical
saturation level and the microphysical properties of the
formation of carbon dioxide cloud particles (Glandorf
et al., 2002), Colaprete and Toon (2003) pointed out
that, since nucleation efficiency decreases when the size
of the nucleating particles decreases, only the
biggest particles are efficiently nucleated and are
able to condensate CO$_2$. They concluded that, as a
result, CO$_2$ clouds on early Mars should contain
few, large particles with average radii greater than
500~$\mu$m. In addition, they showed that a
supersaturation of nearly 35\% is needed to allow
for such an efficient nucleation.
\nocite{Glan:02,Cola:03a}

While the first effect could be well mimicked by
reducing [CCN] as done above,
we have run another set of simulations
to quantify the impact of the
critical supersaturation needed to form clouds in a
cloud free medium.
In these simulations, the number density
of CCN is prescribed as above. However, in a grid cell
where no significant amount of carbon dioxide ice is
present at the beginning of a timestep, CO$_2$ is
allowed to condensate only if the saturation
$s=p/p_{\mbox{sat}}$ reaches a critical value
$s_{\mbox{crit}}=1.35$, i.e. if the temperature in the
cell is below the nucleation temperature
$T_{\mbox{nuc}}(p)=T_c(p/s_{\mbox{crit}})$,
where $T_c(p)$ is the equilibrium
condensation temperature for a CO$_2$ partial pressure
$p$.

The average temperatures reached in these simulations
are also shown in Figure~\ref{fg:nmix} as a function of
[CCN]. As can be seen, the need for supersaturation
decreases the warming effect of CO$_2$ clouds by a few
degrees. However, this variation is weaker than
the one obtained by varying the density of CCN.

\subsection{Impact of obliquity and orbital parameters}
\label{sc:obliquity}

The obliquity and the eccentricity of Mars have strongly varied throughout its
existence. As the evolution of
these parameters is strongly chaotic, it is not possible to
know their values before a few million years ago (Laskar et al. 2004).
Nevertheless, Laskar et al. (2004) showed that the
obliquity could have varied between less than 5$^{\circ}$ and up to more than
60$^{\circ}$ or even $70^{\circ}$. In fact, the average value of the obliquity
over 5~Gyr was estimated to be 37.625$^{\circ}$ with a standard deviation of
13.82$^{\circ}$. The average eccentricity was 0.0690, with standard
deviation 0.0299. More recently, Brasser and Walsh (2011) reanalysed the
stability of the martian obliquity for the Noachian era,
taking into account the fact that before
the late heavy bombardment the giant planets may have been on drastically
different orbits than today, according to Gomes et al. (2005).
For such conditions, they found that the martian obliquity would have remained
chaotic for its most probable mean values, between about 30$^\circ$ and
60$^\circ$, but more stable for the less probable mean
obliquities below 30$^\circ$
(with oscillation amplitude still as high as 20$^\circ$) and
above 60$^\circ$ (with amplitude of 9$^\circ$)

\begin{figure}
  \begin{center}
    \includegraphics[width=8.4cm,angle=-90,clip]{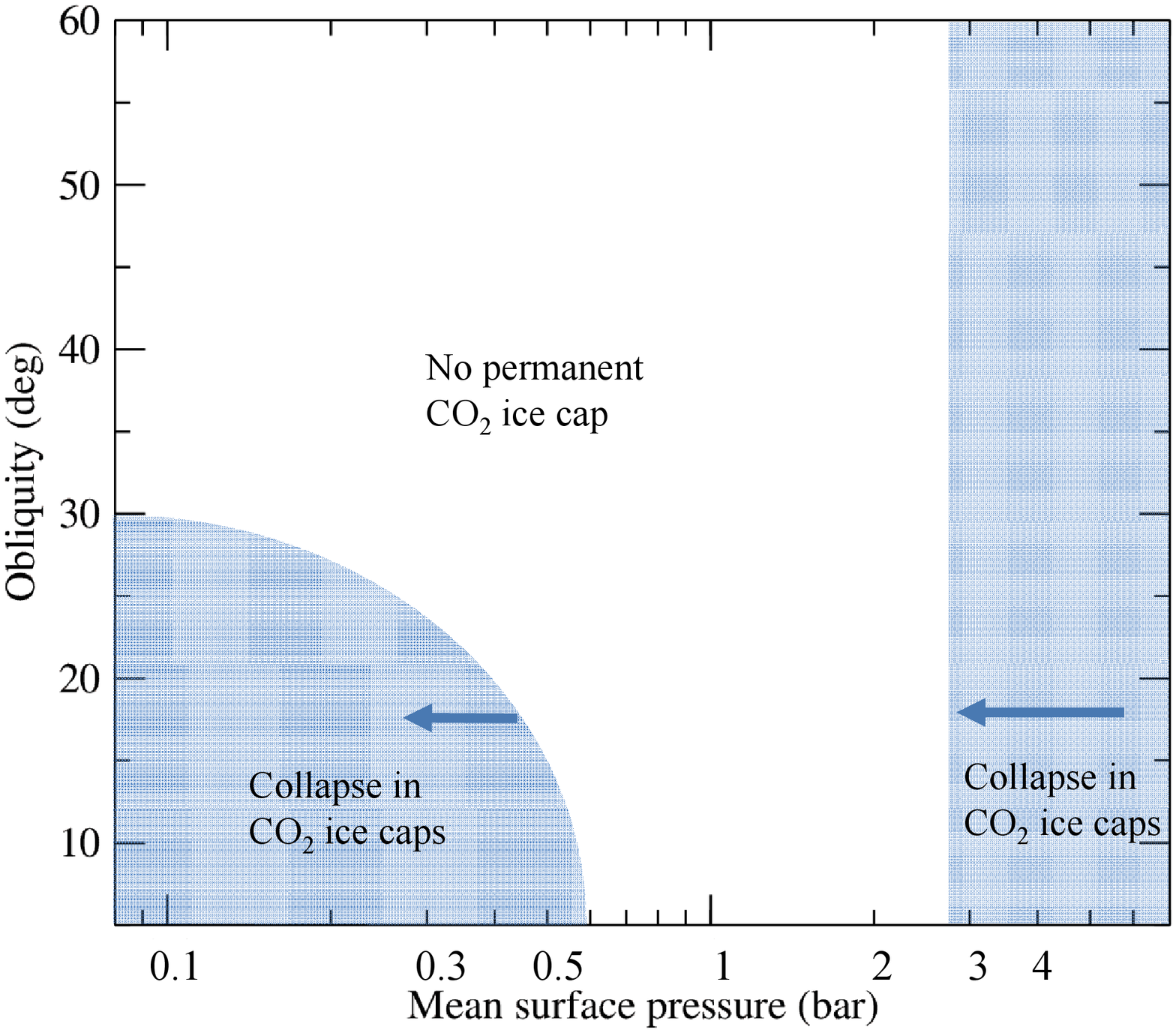}
    \caption{
\label{fg:cap_ob}
Schematic drawing illustrating the range of mean surface
pressures and obliquities for which
permanent CO$_2$ ice caps are predicted to form after 10 years of
simulations. It is based on model runs performed combining pressure
values of 0.1, 0.3, 0.5, 1, 2, 3, 4~bars and obliquities of 10$^{\circ}$,
25$^{\circ}$, 35$^{\circ}$ and 45$^{\circ}$ to explore
the parameters space. The sensitivity to other model parameters is not
shown. In particular,  as in all simulations in this paper,
the CO$_2$ ice albedo and emissivity were set to 0.5 and 0.85
respectively, but one can expect that the figure would be different
assuming other CO$_2$ ice radiative properties. The arrows illustrate
the fact that when permanent CO$_2$ ice forms, part of the atmosphere may
collapse and cause the pressure to decrease below the initial value.  }
  \end{center}
\end{figure}

Varying the obliquity or the season of perihelion in models
of the present-day Mars climate system has a
profound impact on the surface temperatures and the water cycle
(e.g. Haberle et al. 2003, Forget et al. 2006, Montmessin et al. 2007a).
The impact is especially strong at high
latitudes, with the average insolation at the pole proportional to the sine
of the obliquity. The permanent polar caps predicted to form for
$Ps=0.1$ and $Ps=0.3$~bar in the baseline case  (25$^{\circ}$ obliquity)
and discussed in Section~\ref{sc:tsurf} do not form at 35$^{\circ}$ obliquity.
Conversely, we found that a southern permanent polar cap forms when running
with $Ps=$0.5~bar and 10$^{\circ}$ obliquity (whereas there is no such cap in
the baseline case with 25$^{\circ}$ obliquity shown in Figure~\ref{fg:ts_ls}c).
With $Ps=2$~bar, the thick atmosphere prevents the formation of
permanent polar caps even in the $10^{\circ}$ obliquity case. In fact, in that
simulation CO$_2$ does not condense on the surface at all!
Figure~\ref{fg:cap_ob} summarizes the range of mean surface
pressures and obliquities for which
permanent CO$_2$ ice caps
are predicted to form, possibly leadingto
atmospheric collapse.

\nocite{Lask:04,Habe:03,Forg:06,Mont:07cap,Bras:11,Gome:05}

\begin{figure}
  \begin{center}
   \includegraphics[width=8.4cm,angle=-0,clip]{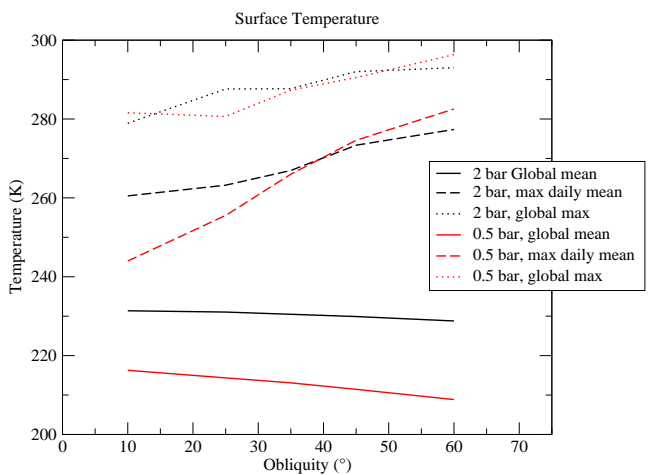}
    \caption{
\label{fg:ts_ob}
Surface temperature statistics (K) as a function of obliquity in the 0.5 and
in the 2~bar mean surface pressure cases (circular orbit).}
  \end{center}
\end{figure}

In general, a higher obliquity corresponds to a stronger seasonal cycle, with
a warmer summer and a colder winter  at mid and high latitudes.
Could this significantly increase
maximum temperatures and allow seasonal melting?
Figure~\ref{fg:ts_ob} shows the global mean, maximum daily mean and planetary
maximum temperatures in the $Ps=0.5$ and 2~bar cases for various obliquities
between 10$^{\circ}$ and 60$^{\circ}$.
In both cases, the global mean temperature slightly
decreases with obliquity.
This is due to the raising of the albedo resulting from  the
seasonal polar caps which are more and more extensive when obliquity increases.
As expected, the maximum daily mean temperature strongly
increases with obliquity, especially in the $Ps=0.5$ bar case for which
temperature above 0$^{\circ}$C are predicted at very high obliquity. Looking at
temperature maps (not shown), we can see that such high temperatures are  reached in both
hemisphere (more in the north), above about 70$^{\circ}$ latitude.

\begin{figure*}
  \begin{center}
   \includegraphics[width=17.cm,angle=-0,clip]{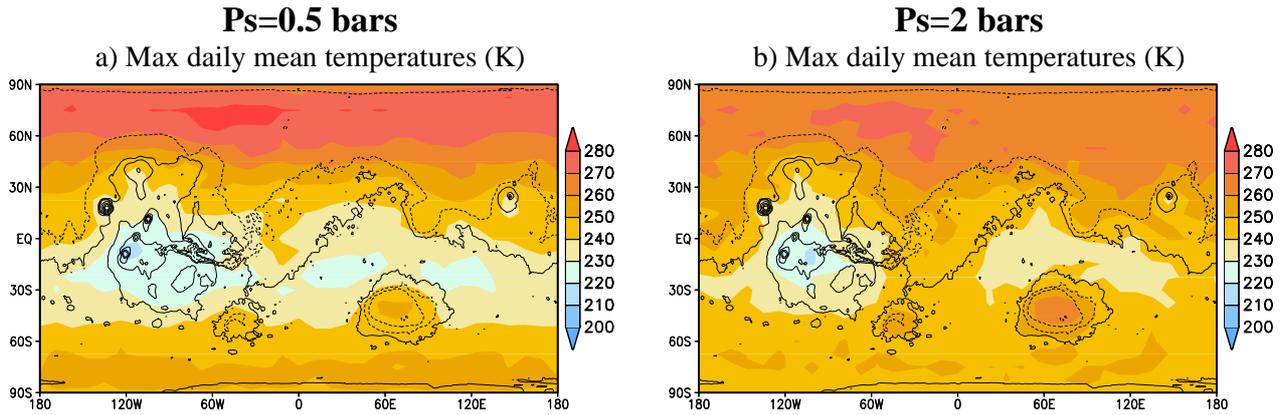}
    \caption{
\label{fg:maptslsp}
Maximum daily mean surface temperatures (K) obtained with simulations with
a 41.8$^{\circ}$ obliquity, eccentricity of 0.1, and Mars closest to the Sun
in Northern summer ($L_s=90^{\circ}$ at perihelion). This illustrates the
warmest daily mean surface temperatures that can be obtained in our simulations
with surface pressure of 0.5 and 2~bars.
    }%
  \end{center}
\end{figure*}

Insolation can be further increased near perihelion
in the case of an eccentric  orbit.
To explore this effect, and in order to create a realistic
``optimum case'' we performed two additional simulations with $Ps=0.5$ and
2~bar, assuming an eccentricity of 0.1, a perihelion during northern summer
solstice,
and an  obliquity set to 41.8$^{\circ}$
(the most likely obliquity for Mars according to Laskar et al. (2004)). The
corresponding maximum daily mean temperatures are shown in
Figure~\ref{fg:maptslsp}. In the 0.5~bar case, seasonal temperatures
above 0$^{\circ}$C are predicted everywhere above~70$^{\circ}$ latitude. If
valley networks and layered deposits were mostly
observed at such high latitudes,
it would be interesting to relate their formation to this type of
orbital configurations. Of course, this is not the case.

\begin{figure}
  \begin{center}
   \includegraphics[width=8.4cm,angle=-0,clip]{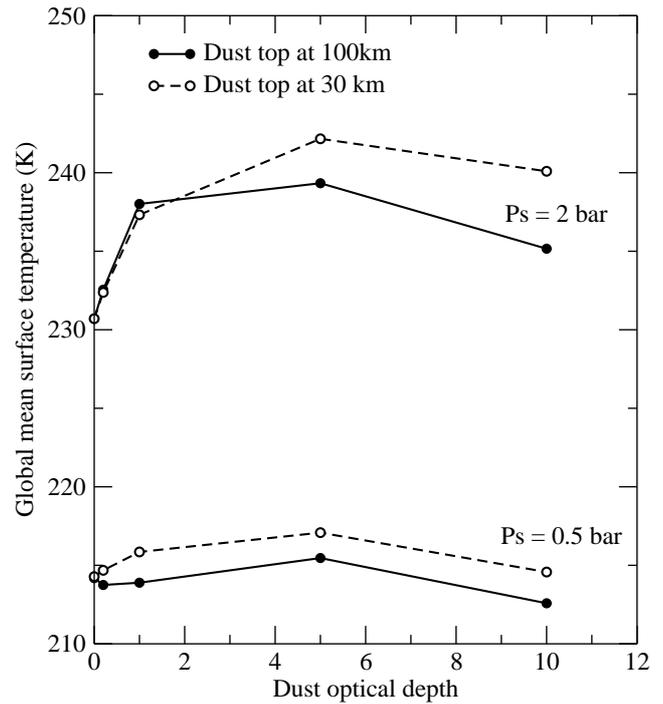}
    \caption{
\label{fg:ts_dust}
Global mean surface temperature (K) as a function of dust opacity
in the 0.5 and in the 2~bar mean surface pressure cases
(circular orbit).  }
  \end{center}
\end{figure}
\begin{figure}
  \begin{center}
   \includegraphics[width=8.4cm,angle=-0,clip]{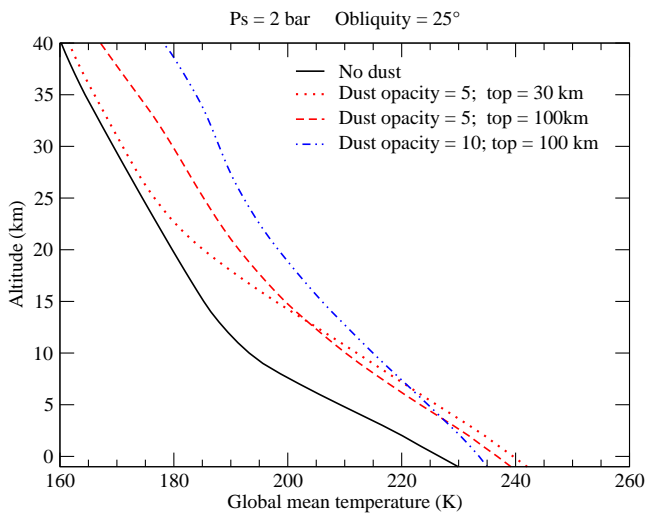}
    \caption{
\label{fg:tprofile_dust}
Global average temperature profile in the 2~bar mean
surface pressure case, and with varying dust loading and vertical
distribution    }
  \end{center}
\end{figure}

\subsection{Impact of atmospheric dust}

\label{sc:dustresults}

To explore the radiative effects of airborne mineral aerosols, we have used the
dust parameterization described in Section~\ref{sc:dustmodel} and performed a
series of simulations with various amount of dust in the
$Ps=$~0.5 and 2~bar cases.
The impact of dust opacity on the global averaged surface
temperature is shown in Figure~\ref{fg:ts_dust}. Two vertical distributions are
employed. When the``dust top'' is set to 100~km, the dust mixing ratio
decreases slowly in the lower 50~km (80\% of the surface value at 50 km)
and more steadily above, up to 100~km. Similarly,
with a dust top at 30~km, most of the dust is confined below 20~km.

In most cases, it is found that dust can
slightly warm the surface. With a dust visible opacity $\tau$ set to $\tau=$5,
surface temperatures are increased by a couple of
Kelvins in the $Ps=$~0.5~bar case, and by more than 10~K with
$Ps=$~2~bar. With dust opacity $\tau=$10, the impact is reduced.
To interpret these results, we can note that airborne dust modifies the
energy balance in four ways:
1) Dust absorbs solar radiation and thus warms the
atmosphere, 2) as a result, it decreases the  CO$_2$ ice clouds opacity
3) Dust decreases the planetary albedo, 4) Dust increases the atmospheric
infrared opacity. The first two effects tend to reduce the atmospheric
greenhouse surface warming and thus the mean surface temperatures. However, the
other two effects work in the opposite direction and actually dominate.
Figure~\ref{fg:tprofile_dust} illustrates the impact of dust on the temperature
profiles in the $Ps=$~2~bar case.
Temperatures  above 10~km are warmed by
15~K with an opacity $\tau=$5 and by more than 20~K with $\tau=$10. This
should decrease the greenhouse effect by increasing the outgoing infrared
radiation, but this is more than compensated by the fact
that the dust infrared opacity allows the mean infrared emission
to take place at higher and colder altitude. The dominant efect is thus an
increase in surface temperature to keep the energy budget in balance.
This is less true with $\tau=$10,
and one can see that the surface temperature is then lower than with $\tau=$5.
In the $\tau=$5 case, the CO$_2$ ice cloud mean opacity is reduced from
about 5 with no dust down to about 1.5 and 0.17 with the
dust top at 30 and 100~km,
respectively.
The colder upper atmosphere temperatures and the thicker clouds explain why
surface temperatures are a few kelvins warmer when the dust is confined in the
lower atmosphere.

\subsection{Role of water}

In this paper we have assumed that Mars was mostly dry, and
neglected the additional greenhouse effect of water vapor.
However, even in the case of a  water-covered Mars, this effect is
limited because our predicted atmospheric temperatures are in most cases so
low that even at saturation the concentration of water vapor remains small.

In the companion paper,  Wordsworth et al. (2012) present simulations
similar to our baseline cases with various CO$_2$ surface pressure,
but in which the greenhouse effect of water vapor
is maximized by assuming 100\% relative humidity everywhere.
As seen on their figures 2 and 3, it is found that even in this overestimated
case, water vapor only increases surface temperatures by a few
kelvins, and thus does not qualitatively change our conclusions.
In reality, water also affects the climate by forming water ice clouds which can
either cool or warm the planet.  This is further discussed in Wordsworth et al. (2012).

\section{Discussion and conclusion}
\label{sc:discussion}

\subsection{Early Mars Climate with a CO$_2$ atmosphere would have been cold.}

In this work, our purpose has been to develop a climate model complete
enough to properly answer the question ``what is the climate on a
Mars-like planet with a thick CO$_2$ atmosphere and a faint young sun?''
Previous work had found contradictory answers to this apparently
simple question (Pollack et al.
1987, Kasting et al. 1991, Forget and Pierrehumbert, 1997, Mischna et al. 2000,
Colaprete and Toon, 2003). By using a complete 3D GCM, revised spectrocopic
properties for collision-induced CO$_2$ , and by systematically
exploring the model sensitivity
to the possible surface pressures, cloud microphysics properties,
obliquity and orbital properties, atmospheric dust loading and model resolution,
we hope to have provided an improved answer to the question.
\nocite{Poll:87,Kast:91,Forg:97,Misc:00}

In our simulations, a wide range of dynamical,
meteorological and climatic phenomenons are observed.
Unlike on present day Mars, and more like on the Earth,
for pressure higher that a fraction of bar,
surface temperatures varies with altitude because of the adiabatic cooling
and warming of the atmosphere when it moves vertically and the increased
coupling between the atmosphere and the surface under higher atmospheric
pressures than today.
In most simulations, CO$_2$ ice clouds cover a major part of the planet
but not all. Their behavior is controlled by a combination of large-scale ascent and descent of air, stationary and travelling waves, and
resolved gravity waves related to the topography. The formation of
CO$_2$ ice seasonal or perennial deposits is also a key process in the climates
that we have explored. In particular, atmospheric
collapse in permanent CO$_2$ ice caps could have buffered the thickness of the
early Mars atmosphere to pressures lower than about 3~bar
(in the unlikely case that enough CO$_2$ was available; see
section~\ref{sc:which_atm}).
Conversely at pressures lower than one bar, we find that if the obliquity is
lower than a threshold value between about $10^{\circ}$ and
$30^{\circ}$ (depending on the initial pressure), a part of the
atmosphere may be trapped in the high latitudes, as on Mars today.
However, as mentioned above,
to properly simulate an equilibrated atmosphere/permanent CO$_2$ ice cap
system, it would also be necessary to take into account the ability of CO$_2$
ice glaciers to flow and spread,  and probably include the effect of slopes and
roughness  (Kreslavsky and Head 2005, 2011).

Our most striking result is that no combination of parameters can yield
surface temperatures consistent with the melting and flow of liquid water as
suggested by the available geological evidence on early Mars.
Previous studies had suggested that CO$_2$ ice clouds could
have strongly warmed the planet thanks to their scattering greenhouse
effect.  Looking at Figures~\ref{fg:ts_ps} and~\ref{fg:nmix},
one can see that the optimum case exhibiting maximum surface temperatures
in our simulations
is with $Ps$=2~bar and [CCN] near 10$^{5}$ - 10$^{6}$~kg$^{-1}$. This actually
corresponds to one of the reference casess that we have studied in
detail.
Even in that case,
the mean cloud warming remains lower than 15~K because of the partial cloud
coverage and the limited cloud optical depth.
We conclude that a CO$_2$ atmosphere could not have raised the
annual mean temperatures above 0$^\circ$C anywhere on the planet.
Summertime diurnal mean surface temperatures above 0$^\circ$C
(a condition which could have allowed rivers and lakes to form)
are predicted for obliquity larger
than 40$^\circ$ at high latitudes but not in locations
where most valley networks or layered sedimentary units are observed.
At most latitudes, above melting temperatures occur,
but only for a few hours during summer afternoons, and assuming
albedo and thermal inertia values which may be unrealistic for snow or ice
deposits. In the companion paper, Wordsworth et al. (2012) present
further results on this subject using  a full water cycle
model that predicts the stability and evolution of surface ice deposits.

\subsection{Was the early sun brighter than expected ?"}
\label{sc:faint}

Like in most previous model studies, we assumed that the early solar flux
was 75\% than today, on the basis of the ``standard solar model''
(e.g. Gough, 1981). Can this model be questioned?
The fundamental reason is very robust: ``The gradual increase in luminosity
during the core hydrogen burning phase of evolution of a star is an inevitable
consequence of Newtonian physics and the functional dependence of the
thermonuclear reaction rates on density, temperature and composition'' (Gough,
1981). In practice, it agrees very well with solar neutrinos and
helioseismology observations. The only way to explain a brighter early sun is to
assume that it was more massive initially and that it subsequently lose the
excess mass in an intense solar wind during its first billion years
(Whitmire et al. 1995). A few percent of the mass
would be sufficient because a star luminosity
strongly depends on its mass. Interestingly, a younger, 
more massive sun has also been suggested recently to explain
inconsistencies between the standard solar model predictions
and recently revised measurements of the solar elemental
abundances
(Guzik and Mussack 2010, Turck-Chieze et al., 2011).  However, observations
of mass loss in young stars do not suggest that they lose significant
mass after the first 0.1 Gyr (Wood et al. 2005, Minton and Malhotra, 2007).
Nonetheless, the faint young sun is still a topic of active investigation.

\nocite{Whit:95,Guzi:10,Turc:11,Wood:05,Mint:07}

\subsection{Other greenhouse gases? }

\label{sc:othergases}


Could supplemental trace greenhouse gases boost the greenhouse
power of a pure CO$_2$ (and H$_2$O) atmosphere? Past studies have notably
explored the effects of
sulfur dioxide (SO$_2$), hydrogen sulfide (H$_2$S), methane (CH$_4$), and
ammonia (NH$_3$).
The key to the success of these gases in solving the faint young
sun paradox for early Mars rests on two main conditions: their ability to absorb
infrared radiation in parts of the spectrum not covered by CO$_2$ and
H$_2$O, and their
ability to sustain the needed concentrations given plausible sources and sinks.

Of the gases listed, NH$_3$ has the potential to plug up the so-called “window”
region (800-1200 cm$^{-1}$)
as it exhibits strong absorption between 700 and 1300~cm$^{-1}$.
This makes it a very powerful greenhouse gas. Kasting et al. (1992)
suggested that 500 ppm of NH$_3$ in a 4-5 bar CO$_2$
atmosphere could raise surface
temperatures to 273 K. However, NH$_3$ is photochemically unstable (Kuhn and
Atreya, 1979) and would require shielding to survive (e.g., Sagan and Chyba
1997; Wolf and Toon, 2010). Methane, which is also unstable and whose photolysis
products could provide the shield, has absorption features at somewhat higher
wavenumbers (1200-1500 cm$^{-1}$), which could further help reduce the outgoing
infrared. However, recent calculations indicate that even at concentrations of
500 ppm CH$_4$ does not significantly boost the greenhouse effect of a pure
CO$_2$/H$_2$O
atmosphere (Jim Kasting, private communication). This is partly the result of an overlap with
water bands, partly because these bands are on the wings of the Planck function,
and partly due to the fact that
CH$_4$ also absorbs in the near infrared, which has
an anti-greenhouse effect. And like NH$_3$, CH$_4$ would require strong sources to
sustain the above concentrations.
\nocite{Kast:92,Kuhn:79nh3,Saga:97,Wolf:10,Rami:12}

As noted in the Introduction, the detection of Noachian-Hesperian aged sulfate
deposits has renewed interest in SO$_2$ and H$_2$S
as potential greenhouse gases. An
obvious source for these gases is volcanic activity, which was much higher in
the past than it is today. H$_2$S has absorption features in the window
region but they are weak,
and its stronger bands at lower wavenumbers occur at the
tail end of the Planck function. On the other hand, SO$_2$ $\nu_2$
fundamental near
520 cm$^{-1}$ is near the peak of a 273 K Planck function, and it also has
significant opacity at the high wavenumber end of the window region. Thus,
SO$_2$
has received most of the attention (e.g., Halevy et al., 2007; Johnson et al.,
2008).

To be effective, SO$_2$ needs to build up to concentrations around the 10 ppm level
or higher in moderately thick atmospheres (surface pressures greater than 0.5
bar). In the model of Johnson et al. (2008) global mean surface temperatures
well above freezing are achieved with SO$_2$ concentrations in this range. This is
the only modern model to predict such warm conditions with a pure gaseous
greenhouse effect. However, in practice the SO$_2$ greenhouse faces significant
challenges as SO$_2$ readily converts to aerosols regardless of the redox state of
the atmosphere. And as on Earth, these aerosols should have a net cooling effect on
surface temperatures (Tian et al., 2010). Furthermore, SO$_2$ is highly soluble
and will washout quickly when conditions become warm enough for rainfall.

Thus, while the supplemental greenhouse gases considered thus far in the
literature cannot be ruled out, they require special circumstances to play a
significant role in warming the surface of Mars to the melting point of liquid
water. To date, no model has been published that convincingly demonstrates a
gaseous greenhouse can solve the faint young sun problem for early Mars.
Within that context, using our Global Climate Model, we plan to explore the
radiative effect and the chemical cycle of these gases, starting with the sulfur
cycle.

\nocite{Tian:10,Hale:07,John:08}

\subsection{A cold early Mars warmed episodically and locally?}

The idea that Early Mars climate may have been
cold is not new. Several studies have suggested that
the geomorphological evidence related to liquid water
could be explained by assuming hydrothermal convection and the formation of
aquifers driven by geothermal heat and heat associated with impacts
(Squyres and Kasting, 1994, Gaidos and Marion, 2003). Erosion and surface
alteration may have occurred by saline and acidic liquid solutions
(brines) which can be stable at sub-zero temperatures (Fairen, 2010).
Impacts may have induced melting and rainfall on a global scale (Toon et al.
2010) or locally  (e.g. Mangold, 2012, and reference therein)  which may
explain the formation of fluvial valleys even under cold background climate
conditions. Similarly, while clay minerals have been thought to be
an indication of surface weathering by liquid water and thus an
indication of warm and wet past conditions, Ehlmann et al. (2011)
showed that, in many locations, available data instead indicates clay formation
by hydrothermal groundwater circulation possibly
operating in cold, arid surface conditions. Furthermore,
Meunier et al. (2012) proposed that the iron-and-magnesium-rich clay
formed direcly by the precipitation from water-rich magama-derived fluids.

This new paradigm is not without problems. In particular, it is more and more
clear that ``early'' Mars encompasses several epochs.
While a cold early planet scenario
may be consistent with relatively recent late Noachian or
Hesperian geological features such as delta fans or the most recent fluvial
features, several authors have argued that it could not explain
the morphology of large valley networks, the very high global
erosion rate inferred for the early Noachian period, or the formation of
extensive deposits of phyllosilicates like in the Mawrth Vallis region
(e.g. Craddock and Howard, 2002, Bibring et al. 2006, McKeown et al. 2009).

In any case, there is no doubt that conditions on Mars during the first
billion years of its history were dramatically different from those at
present. A thicker atmosphere was probably necessary to allow
 water to flow for long distances across the surface.
Our simulations of the resulting climate provide clues on the environment, and
notably show that with surface pressure
above a fraction of a bar, the southern highlands become cold traps where ice
may migrate and replenish water sources. This hydrological cycle is
modeled and discussed in detail in the companion paper
(Wordsworth et al., 2012).

\nocite{Squy:94,Gaid:03,Fair:10,Toon:10,Mang:12,Ehlm:11,Crad:02,Bibr:06,McKe:09}
\nocite{Meun:12}

\section*{Acknowledgments}
We are grateful for the comments on the text and scientific suggestions
provided by James W. Head.
We thank N. Mangold, B. Charnay, F. Leblanc, A. Spiga, J-P. Bibring
for helpful discussions.  We also appreciate the reviews provided by
E. Kite and S. Clifford.

\end{document}